\documentclass[a4paper,11pt,twoside,openright]{article}

\usepackage[utf8]{inputenc}                         % Ecrire avec les conventions du français.
\usepackage[dvips,pdftex]{graphicx}%\usepackage{graphicx}
\usepackage[left=2.5cm,top=3cm,right=2.5cm,bottom=3cm]{geometry}    % Modification des marges
\usepackage{amsmath,amsthm,amssymb}                             % Utilisation de certains packages de AMS
\usepackage{fancyhdr}
\usepackage{color}

\usepackage{url}

\title{The obliquity of Enceladus}
\author{Rose-Marie Baland$^{a,b}$, Marie Yseboodt$^b$, Tim Van Hoolst$^b$\\
\textit{$^a$ Universit\'{e} catholique de Louvain, Earth and Life Institute, Louvain-la-Neuve, Belgium.}\\
\textit{$^b$Royal Observatory of Belgium, Ringlaan 3, B-1180 Brussels, Belgium.}\\Email: Rose-Marie.Baland@oma.be}
\date{December 2015\\ Paper accepted for publication in Icarus}

\begin{document}

   \maketitle

\tableofcontents
\newpage

\section*{Abstract}

The extraordinary activity at Enceladus' warm south pole indicates the presence of an internal global or local reservoir of liquid water beneath the surface. 
While Tyler (2009, 2011) has suggested that the geological activity and the large heat flow of Enceladus could result from tidal heating triggered by a large obliquity of at least $0.05^\circ-0.1^\circ$, theoretical models of the Cassini state predict the obliquity to be two to three orders of magnitude smaller for an entirely solid and rigid Enceladus. We investigate the influence of an internal subsurface ocean and of tidal deformations of the solid layers on the obliquity of Enceladus. Our Cassini state model takes into account the external torque exerted by Saturn on each layer of the satellite and the internal gravitational and pressure torques induced by the presence of the liquid layer. As a new feature, our model also includes additional torques that arise because of the periodic tides experienced by the satellite. We find that the upper limit for the obliquity of a solid Enceladus is $4.5 \times 10^{-4}$ degrees and is negligibly affected by elastic deformations. \textcolor{black}{The presence of an internal ocean decreases this upper limit by $13.1\%$, elasticity attenuating this decrease by only $0.5\%$}. For larger satellites, such as Titan, elastic effects could be more significant because of their larger tidal deformations. \textcolor{black}{As a consequence, it appears that it is easier to reconcile the theoretical estimates of Titan's obliquity with the measured obliquity than reported in previous studies wherein the solid layers or the entire satellite were assumed to be rigid.} Since the obliquity of Enceladus cannot reach Tyler's requirement, obliquity tides are unlikely to be the source of the large heat flow of Enceladus. More likely, the geological activity at Enceladus' south pole results from eccentricity tides. Even in the most favorable case, the upper limit for the obliquity of Enceladus corresponds to about two meters at most at the surface of Enceladus. This is well below the resolution of Cassini images. Control point calculations cannot be used to detect the obliquity of Enceladus, let alone to constrain its interior from an obliquity measurement. 

\newpage
\section{Introduction}

Enceladus is a small (504 km in diameter) and very bright (visual geometric albedo of 1.4) satellite of Saturn which presents extraordinary activity at its south pole, more precisely at the location of the long fractures known as the ``tiger stripes''. Plumes of water vapor and ice particles rising up to more than 200 km above the surface have been observed from the images taken by the Cassini spacecraft \textcolor{black}{(Porco et al., 2006; Spencer et al., 2009)}. These observations, complemented by thermal measurements indicating an anomalously warm south pole region \textcolor{black}{(Spencer et al., 2006)}, have been interpreted as possible evidence of a liquid water reservoir beneath the surface (Porco et al., 2006). \textcolor{black}{Moreover, this liquid reservoir has been in prolonged contact with an underlying rocky
core, as suggested by the observation of salty ejected ice grains in the plume (Postberg et al., 2011).} The nature of the reservoir (global ocean or local reservoir) is still debated. \textcolor{black}{According to Iess et al. (2014), gravity, topography, and heat flux measurements are in favor of a local reservoir, although the gravity measurements cannot rule out a global ocean. Thermal models are in general not consistent with the global ocean case (e.g. Tobie et al., 2008 and Roberts and Nimmo 2008). However, McKinnon (2015), revising the analysis of Iess et al. (2014) by taking into account the rapid rotation of the moon, concludes that the gravity and topography data actually imply a global ocean.} \textcolor{black}{Patthoff and Kattenhorn (2011) have interpreted the fracture patterns in the south pole region as the result of long-term (timescales of tens of thousands to millions of years) nonsynchronous rotation of a floating ice shell above a global liquid ocean. Measurements of rotation features 
that are observable over shorter time scales (e.g. the duration of a space mission) may help to discriminate between the global and local hypotheses, since a global ocean may indeed decouple} the surface from the interior and lead to a different rotation state. 

Using images of surface control points, Porco et al. (2006) detected no librations (variations in the rotation rate) \textcolor{black}{which, given the uncertainties in the solution, imply an upper limit for the libration amplitude of $1.5^\circ$ ($6.6$ km at the surface)}. \textcolor{black}{Giese (2011) claimed the detection of $0.056^\circ$ diurnal librations after a fit of control points measurements to the libration model of Rambaux (2010) for a solid Enceladus. In contrast to Giese, Thomas et al. (2016), using a similar approach, have found an amplitude for the forced diurnal libration of $0.12^\circ$ (523 m), too large to be consistent with a solid Enceladus, thereby suggesting the presence of a decoupling internal global ocean.}

Another component of the rotation is the spin precession, characterized by the \textcolor{black}{obliquity which is the angle from the orbit plane normal to the spin axis}. For a synchronous triaxial satellite, it is often assumed that the obliquity is that of the Cassini state (an equilibrium rotation state where the precession of the spin axis follows the precession of the orbit axis). When the orbital precession is uniform and the satellite is solid and rigid, the obliquity $\varepsilon$ of the Cassini state is constant and given, at first order in the small angles $\varepsilon$ and $i$, by (e.g. Noyelles 2010, Appendix B)
\begin{eqnarray}\label{solidrigid}
 \varepsilon&=&-\frac{2i C \dot\Omega}{3M_eR^2(-C_{20}+2 C_{22})n+2C\dot\Omega},
\end{eqnarray}
where $\dot\Omega$ is the constant orbital precession rate, $i$ is the orbital inclination (constant if the precession is uniform), $n$ is the mean motion, $M_e$ and $R$ are the mass and the mean radius, $C_{20}$ and $C_{22}$ are the second-degree gravity field coefficients, and $C$ is the polar moment of inertia.

From control points calculations using Cassini's images, Giese (2014) claimed that the obliquity of Enceladus is limited to less than $0.05^\circ$ at the $2\sigma$ level. However, he has assumed the spin axis to be fixed in space during the observation period 2005-2012, instead of being in the Cassini state, \textcolor{black}{even though the orbital precession period is 2.4 years}. This is unlikely since the system is expected to be damped and without recent punctual excitation. The assumption of Giese leads to determine the mean orientation of the orbit pole (which is the Laplace pole) instead of the spin pole orientation, and its obliquity estimate is in fact an estimate of the mean orbital inclination \textcolor{black}{(which is $0.008^\circ$, actually smaller than the upper limit of $0.05^\circ$)}. 

A theoretical published model for a solid and rigid Enceladus in the Cassini state demonstrated a two to three order of magnitude smaller upper limit of $0.0015^\circ$ (Chen and Nimmo, 2011). The solution of Chen and Nimmo (2011) is equivalent to the solution given in Eq.  (\ref{solidrigid}), \textcolor{black}{with the difference that their estimate of the obliquity has been performed using reasonable estimates of $C_{20}$ and $C_{22}$, both of which were still undetermined at the time.} From the measured gravity coefficients $C_{20}=-5435.2\times 10^{-6}$ and $C_{22}=1549.8\times 10^{-6}$ and the estimate of the normalized mean moment of inertia $I/M_eR^2=0.335$ of Iess et al. (2014) taken as a first order approximation for the normalized polar moment of inertia $C/M_eR^2$, and from the mean precession rate $\dot \Omega=-2.65761$ rad/yr and mean inclination $i=0.00819^\circ$ derived from JPL/SAT375 ephemeris\footnote{\url{http://naif.jpl.nasa.gov/pub/naif/generic_kernels/spk/satellites/sat375.cmt}}, the rigid solid uniform obliquity $\varepsilon$ is $0.00036^\circ$, which is about four times smaller than Chen and Nimmo's estimate.
 
In this paper, we propose a new theoretical model for the Cassini state, taking into account that Enceladus experiences a nonuniform orbital precession (see Bills, 2005 for an application to the Galilean satellites), may harbor an internal global ocean (see Baland et al., 2011, for the first application of such a model to Titan), and experiences periodic elastic deformations. To our knowledge, a theoretical model for the spin precession of a synchronous satellite under the influence of periodic tides does not exist yet, contrary to the case of the librations (Van Hoolst et al., 2013; Richard et al., 2014; Jara-Oru\'{e} and Vermeersen, 2014). With this new model, we will assess the possibility of a resonant amplification of the solution (due to the non uniformity of the orbital precession and/or the presence of the internal global ocean and/or the elastic deformations of the solid layers). Such a resonance may lead to a high value of the obliquity, as already demonstrated theoretically for Titan and the Galilean satellites (Baland et al., 2011, 2012).

Obliquity tides are often proposed as a possible cause for the geological activity of icy satellites. The need for a new Cassini state model comes mainly from the hypothesis made by Tyler (2009, 2011) that the jets of Enceladus could be associated with a strong heat flux resulting from tidal heating triggered by a high obliquity and the associated large scale flow in the ocean. According to Tyler, to account for the observed flux, the obliquity of Enceladus must have a minimal value of $0.05^\circ-0.1^\circ$. Similarly, Nimmo and Spencer (2015) argue that Triton likely harbors an internal subsurface ocean and that the resurfacing of the satellite is due to convection driven by obliquity tides in the outer ice shell. Rhoden et al. (2015) linked the plume activity recently detected at Europa's south pole to the stress variations induced by obliquity tides in an outer thin elastic ice shell mechanically decoupled from the interior by a subsurface ocean. In addition, the new Cassini state model could benefit the study of the rotation of other synchronous satellites such as Titan or Ganymede. Titan's obliquity has been measured (Stiles et al., 2008, 2010) and the tidal deformations may change the interpretation in terms of internal structure done in Baland et al. (2014). The determination of the rotation state of Ganymede is one of the goals of the future JUICE mission \textcolor{black}{(Grasset et al., 2013)}, and an accurate model for the obliquity will be needed to interpret future measurements.

The plan of the paper is as follows. In section 2, we extend the solid Cassini state model to the case of an elastic satellite with periodic tidal deformations. \textcolor{black}{We compare our solution with the solution derived by Chen and Nimmo (2011) using a theoretical model for a solid and rigid satellite,} which is equivalent to the solution given by Eq.  (\ref{solidrigid}). In section 3, we consider the case where Enceladus harbors a global internal ocean, and where gravitational and pressure torques arise between the solid layers. We compare the cases with and without elastic deformations to each other and to the respective entirely solid cases. In section 4, we determine a new theoretical estimate of the upper limit for Enceladus obliquity, and compare it to Tyler's requirement. We also discuss the influence on the results of deviations of Enceladus' shape from the hydrostatic equilibrium and the implications of elasticity for the obliquity of other satellites such as Titan. We present concluding remarks in section 5. 

\section{The Cassini state for an entirely solid Enceladus}
\label{section2}

\subsection{Angular momentum equation}

\subsubsection{Rigid case}

We first consider that Enceladus is entirely solid and rigid. The spin precession can be considered as a long-term behavior, compared to the orbital revolution. The angular momentum equation governing the spin precession, averaged over the short orbital/diurnal period, correct up to the first order in obliquity, eccentricity, and orbital inclination, is given by (Bills, 2005, corrected for a sign error in Baland et al., 2011)
\begin{eqnarray}\label{rigidcase}
 n C \frac{d\hat s}{dt}&=&n \kappa (\hat s \wedge \hat n),\\
\kappa&=&\frac{3}{2}M_eR^2(-C_{20}+2 C_{22})n=\frac{3}{2}(C-A)n,
\end{eqnarray}
where $\hat s=(s_x,s_y,s_z\simeq 1)$ and $\hat n=(n_x,n_y,n_z\simeq 1)$ are the unit vectors along the spin axis and the normal to the orbit, expressed in Cartesian coordinates $(x,y)$ of an inertial plane and $z$ along the normal to it. The inertial plane is taken here as the Laplace plane, which is the mean orbital plane of Enceladus. $A$ is the smallest principal moment of inertia. The left-hand member of Eq. (\ref{rigidcase}) is the variation of the angular momentum $\vec L=n C\hat s$ while the right-hand member is the gravitational torque (averaged over the orbital period) exerted by Saturn. The strength of the coupling, $\kappa$, depends on the moment of inertia difference $(C-A)$. 

Enceladus is not in hydrostatic equilibrium, so this moment of inertia difference can be written as the sum of an hydrostatic part and of a non-hydrostatic one:
\begin{equation}
 (C-A)=(C-A)_{HE}+(C-A)_{NHE}.
\end{equation}
The hydrostatic part results from the static deformations of Enceladus under the influence of its own centrifugal potential and of the tidal potential raised by Saturn, and can be written as
\begin{equation}
 (C-A)_{HE}=(C-A)_{c}+(C-A)_{t}=\frac{1}{3} k_f (q_r-q_t) M_eR^2,
\end{equation}
with $k_f$ the fluid Love number, $q_r=n^2R^3/GM_e$ the ratio of the centrifugal acceleration to the gravitational acceleration and $q_t=-3\frac{GM_p}{GM_e}\left(\frac{R}{a}\right)^3$ the tidal parameter. $M_{p}$ is the mass of the planet Saturn, $G$ is the universal gravitational constant, $a$ is the semi-major axis of Enceladus. \textcolor{black}{Note that due to Kepler's third law $GM_p=n^2a^3$ and hence $q_t=-3q_r$.}

\subsubsection{Elastic case}

In the elastic case, the periodic tidal bulge presents a component \textcolor{black}{due to} the obliquity tide and that depends on the difference between the spin and orbital precessions. The other components of the periodic tidal bulge are the radial and librational tides, which play no role here, because the effect of the first is of second order and the effect of the latter vanishes when the angular momentum equations are averaged over the short diurnal period (see Appendix A). An effect of the periodic obliquity bulge is to decrease the torque exerted by Saturn by decreasing the \textcolor{black}{hydrostatic} part of the moment of inertia difference due to the tidal potential by a factor $\frac{k_f-k_2}{k_f}$, with $k_2$ the tidal Love number of Enceladus. The total effective moments of inertia difference is:
\begin{eqnarray}
 (C-A)^{\textrm{eff}}&=&(C-A)_{NHE}+\frac{1}{3} k_f \left[q_r-q_t \left(\frac{k_f-k_2}{k_f}\right) \right] M_eR^2,\\
\label{7} &=&(C-A)-k_2 M_e R^2 q_r.
\end{eqnarray}
The obliquity bulge also affects the angular momentum which is now expressed as (see Appendix A):
\begin{eqnarray}
 \vec L&=&n \tilde C\hat s -n(\tilde C- C)\hat n,\\
\label{9} \tilde C &=&\, C -\frac{1}{2}k_2 M_e R^2 q_r.
\end{eqnarray}

\textcolor{black}{By introducing the effects of the periodic obliquity bulge on the angular momentum and torque, the angular momentum equation can be written as}
\begin{eqnarray}\label{elasticcase}
 n\, \tilde C\frac{d \hat s}{dt}&=&n \kappa_{el} (\hat s \wedge \hat n)+n(\tilde C- C)\frac{d \hat n}{dt},\\
\label{kappael} \kappa_{el}&=&\frac{3}{2} n(C-A)^{\textrm{eff}} =\frac{3}{2} n \left\lbrace(C-A)-k_2 M_e R^2 q_r\right\rbrace,
\end{eqnarray} 
where $\kappa_{el}$ is the coupling strength for the \textcolor{black}{case of a solid and elastic moon}. For a $k_2$ equal to $k_f$ (fictive case of an entirely fluid satellite), the torque would not vanish contrary to the case of elastic librations (Van Hoolst et al., 2013), because the difference $(C-A)$ has a part due to the response to the centrifugal potential, while the difference between the equatorial moments of inertia, $(B-A)$, governing the librations in longitude, is only due to the tidal potential.

\textcolor{black}{The relative effect of elasticity on the angular momentum is two orders of magnitude lower than its relative effect on the torque, since the polar moment of inertia $C$ is two orders of magnitude larger than the moments of inertia difference $(C-A)$, while the substracted factor ($\propto k_2 M_e R^2 q_r$) is of the same order (see Eqs. (\ref{9}) and (\ref{kappael})).}

\subsection{Free mode}

\textcolor{black}{We first average the orbital precession, in order to get the free mode of the angular momentum equation (free precession). We set $\hat n$ equal to $(0,0,1)$  in Eqs. (\ref{rigidcase}) or (\ref{elasticcase}) and we obtain a solution of the form
\begin{eqnarray}
 \hat s_{f}&=&(\Upsilon_{f} \cos(\omega_f t+\phi_f), \Upsilon_{f} \sin(\omega_f t+\phi_f),1 ),
\end{eqnarray}
where $\omega_{f}$ is the free precession frequency. A free precession can be triggered by unusual events (e.g. impact) and the amplitude $\Upsilon_{f}$ and phase $\phi_f$ would depend on initial conditions. However, dissipation processes due to tides damp the free precession and the amplitude $\Upsilon_{f}$ can be assumed to be zero. The free frequency is still of interest, though, since it is involved in the forced solution (see next section).}

\textcolor{black}{In the rigid case, $\omega_{f}$ depends only on the distribution of mass inside Enceladus and is given by the ratio of the external coupling strength over the polar moment of inertia, which represents the resistance to the forcing,
\begin{equation}
 \label{13}\omega_{f}=\frac{\kappa}{C}=\frac{3}{2}n\,\frac{(C-A)}{C}
\end{equation}}

\textcolor{black}{In the elastic case, the free frequency $\omega_{f}^{el}$  is given by
\begin{equation}
 \omega_{f}^{el}=\frac{\kappa_{el}}{\tilde C}=\frac{3}{2}n\,\frac{(C-A)-k_2 M_e R^2 q_r}{C-\frac{1}{2}k_2 M_e R^2 q_r}.
\end{equation}
By decreasing the coupling strength $\kappa_{el}$ with respect to its rigid counterpart $\kappa$, elasticity also decreases the free frequency $\omega_{f}^{el}$ with respect to $\omega_{f}$. The small decrease of $\tilde C$ with respect to $C$ is not expected to have a noticeable influence on $\kappa_{el}$.}

\subsection{Forced solution for a uniform orbital precession}
\label{section23}
\textcolor{black}{When the orbital precession rate with respect to the Laplace plane $\dot \Omega$ is constant, the orbital precession is uniform and the spin axis precesses at the same constant rate and remains coplanar with the normal to the orbit and the normal to the Laplace plane, while the obliquity is a constant angle (see e.g. Henrard, 2005). Therefore, the unit vectors $\hat n$ and $\hat s$ can be written, at first order in the small angles $i$ and $\varepsilon$, as (see Bills, 2005 and Baland et al., 2011)}
\begin{eqnarray}
\label{15} \hat n&=&(i \cos(\Omega-\pi/2),i\sin(\Omega-\pi/2),1 ),\\
\label{16}  \hat s&=&( (i+\varepsilon) \cos(\Omega-\pi/2),(i+\varepsilon)\sin(\Omega-\pi/2),1 ),
\end{eqnarray}
\textcolor{black}{with $\Omega$ and $i$ the node longitude and inclination of the orbit with respect to the Laplace plane, respectively (see Fig. \ref{FigA1}, where, as a result of the coplanarity, the node longitude of the spin axis $\psi$ is equal to $\Omega$).} 

\textcolor{black}{The expression of $\varepsilon$ (see Eq. (\ref{solidrigid})), the constant obliquity of a solid and rigid satellite, is found by introducing Eqs. (\ref{15})-(\ref{16}) in the angular momentum equation (\ref{rigidcase}). In terms of the free mode frequency $\omega_f$ (Eq. (\ref{13})), $\varepsilon$ can be written as}
\begin{equation}
 \label{eq10}\varepsilon=-\frac{i \dot\Omega}{\omega_{f}+\dot\Omega}.
\end{equation} 

\textcolor{black}{The constant obliquity of a solid and elastic satellite, $\varepsilon_{el}$, is obtained by introducing Eqs. (\ref{15})-(\ref{16}) in Eq. (\ref{elasticcase}). In terms of the free mode frequency $\omega_f^{el}$, $\varepsilon_{el}$ can be written similarly as its rigid counterpart (\ref{eq10}), but with a multiplying factor $C/\tilde C$:}
\begin{equation}\label{solidelastic}
 \varepsilon_{el}=-\frac{C}{\tilde C}\frac{i \dot\Omega}{\omega_{f}^{el}+\dot\Omega}.
\end{equation} 

\subsection{Forced solution for a nonuniform orbital precession}

\subsubsection{Rigid case}

\textcolor{black}{When the orbital precession rate $\dot \Omega$ is not constant, the spin axis, the normal to the orbit, and the normal to the Laplace plane are not coplanar while the orbital inclination and the obliquity vary with time. The case of the spin precession of a solid and rigid satellite due to a nonuniform orbital precession has been explicitly described in Bills (2005). By expressing the orbital precession as a series of surperimposed uniform orbital precessions, he showed that the coplanarity is restored on a mode-by-mode basis.}

\textcolor{black}{First, the vectorial equation (\ref{rigidcase}) is projected on the Laplace plane, 
\begin{equation}\label{rigidprojected}
 \frac{d S}{dt}=I \frac{\kappa}{C} (N-S),
\end{equation}
with $S=s_x+I s_y$ and $N=n_x+I n_y$, $I=\sqrt{-1}$. This expression is correct up to the first order in the small time-varying obliquity $\varepsilon(t)$ and orbital inclination $i(t)$ that are given by 
\begin{equation}\label{incl}
i(t)\simeq\|N\|\quad\textrm{and}\quad \varepsilon(t)\simeq\|S-N\|,
\end{equation}
since $(s_x,s_y)\simeq((i(t)+\varepsilon(t)) \cos(\Omega-\pi/2),(i(t)+\varepsilon(t))\sin(\Omega-\pi/2))$ and $(n_x,n_y)\simeq(i(t) \cos(\Omega-\pi/2),i(t)\sin(\Omega-\pi/2))$.}

\textcolor{black}{Then, the orbital precession is written as a series expansion
\begin{equation}\label{precession}
N=\sum_j i_j\, e^{I (\dot\Omega_j t+ \gamma_j-\pi/2)},
\end{equation}
with $i_j$ the inclination amplitudes associated with the orbital node precession frequencies $\dot\Omega_j$ and the phases $\gamma_j$.}

\textcolor{black}{Finally, the forced spin precession, which is the forced solution of Eq. (\ref{rigidprojected}), is also written as a series expansion where the precession rates and phases are the same as in $N$, as a result of the mode-by-mode coplanarity:}
\textcolor{black}{\begin{equation}\label{forcedsol}
S=\sum_j (i_j+\varepsilon_j)\, e^{I (\dot\Omega_j t+ \gamma_j-\pi/2)},
\end{equation}
where, correct up to the first order in $\varepsilon_j$ and $i_j$, the obliquity amplitudes $\varepsilon_j$ associated with the frequencies $\dot\Omega_j$ are given by
\begin{equation}\label{rigidfirstorder}
\varepsilon_j=-\frac{i_j\dot\Omega_j}{\omega_f+\dot\Omega_j}.
\end{equation}}

\subsubsection{Elastic case}

\textcolor{black}{In the elastic case, the vectorial equation (\ref{elasticcase}) projected on the Laplace plane reads
\begin{equation}
 \label{26}\frac{d S}{dt}=I \frac{\kappa_{el}}{\tilde C} (N-S)+\frac{\tilde C-C}{\tilde C}\frac{d N}{dt}.
\end{equation}
With a solution of the form
\begin{equation}\label{forcedelsoleals}
S=\sum_j (i_j+\varepsilon_j^{el})\, e^{I (\dot\Omega_j t+ \gamma_j-\pi/2)},
\end{equation}
the obliquity amplitudes of the elastic satellite $\varepsilon^{el}_{j}$ are given by
\begin{equation}\label{elasticfirstorder}
\varepsilon^{el}_{j}=-\frac{C}{\tilde C}\frac{i_j\dot\Omega_j}{(\omega_{f}^{el}+\dot\Omega_j)}.
\end{equation} 
The actual time-varying obliquity of the elastic body $\varepsilon_{el}(t)$ is defined similarly as in Eq. (\ref{incl}).}

\subsection{Results}

As mentioned in the introduction, using the measured gravity coefficients $C_{20}=-5435.2\times 10^{-6}$ and $C_{22}=1549.8\times 10^{-6}$ and the estimate of $C/M_eR^2\simeq I/M_eR^2=0.335$ of Iess et al. (2014), the uniform solution has a constant obliquity of $0.00036^\circ$ in the rigid case. In the nonuniform rigid case, Chen and Nimmo (2011) concluded that the time-varying obliquity would vary by 20\% with respect to its small mean value. They used a secular variation model that includes the perturbations from the other satellites of Saturn to generate a series for the orbital precession. They also used reasonable estimates of the gravity field coefficients, not measured at that time. Here, we examine their conclusion, using the measured gravity field of Iess et al. (2014) and a different approach for the modeling of the orbital precession. 

\begin{table}[!htb] 
% \scriptsize  
\centering          
\begin{tabular}{c c c c c c c }   
\hline   $j$&$i_j$&$\dot\Omega_j$ &$T_{j}$ &$\gamma_j$ & $\varepsilon_j$ & $\varepsilon_j$ \\
   & (deg)& (rad/year)&(years)& (deg) &  ($10^{-4}$deg) & (m)\\ 
\hline                    
      $1$&$0.00819$&$-2.6576$&$-2.364$&$98.5$   & $3.547$ & $1.561$\\
      $2$&$0.00420$&$-1.2612$&$-4.982$&$15.3$ & $0.844$&  $0.371$\\
      $3$&$0.00017$&$0.4268$&$14.721$&$272.7$  & $-0.011$& $-0.005$\\
      $4$&$0.00013$&$ -6.3715$&$-0.986$&$108.5$  & $0.144$& $0.063$\\
      $5$&$0.00010$&$-0.1754$&$-35.813$&$103.$  & $0.003$& $0.001$\\
      
\hline
\end{tabular}
\caption{{\label{ephem}}Columns 2 to 5: amplitudes, frequencies, periods, and phases of the orbital precession of Enceladus with respect to the Laplace plane of Enceladus and with respect to J2000, based on the quasi-periodic decomposition of the JPL/SAT375 ephemeris over the interval 1900-2100. The secular trend due to the orbital precession of Saturn has been removed.  The Laplace plane has a node (which defines the x-axis) of $244.5^\circ$ and an inclination of $0.00107^\circ$ with respect to the equatorial plane of Saturn at J2000. The obliquity amplitudes of the \textcolor{black}{Cassini state of a solid and rigid Enceladus ($\varepsilon_j$)} are given in the last two columns, in degrees and meters on the surface.}
\end{table}

\begin{figure}[!htb]
\begin{center}
\includegraphics[width=6cm]{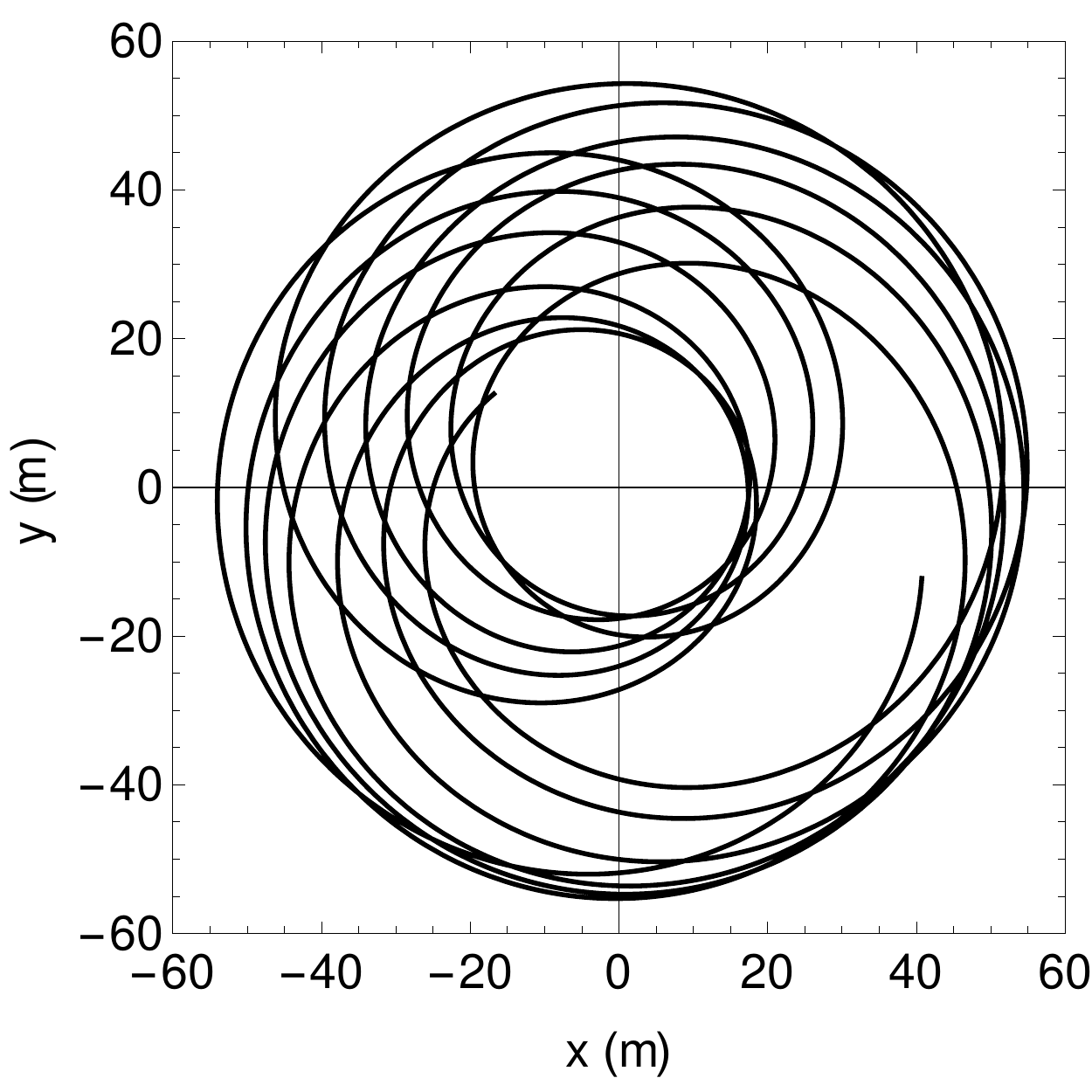}
\caption{\label{orbfig}Projection of the normal to the orbit $N$ (multiplied by Enceladus' radius $R$) on the Laplace plane over 30 years (\textcolor{black}{$\cong 13$} times the main orbital precession period), beginning on \textcolor{black}{J2000}. The amplitude of the orbital precession (i.e. the inclination) corresponds to about $60$ m at the level of Enceladus' surface, at best. The orbital precession is essentially a superposition of the first and second terms of the series of Tab (\ref{ephem}).}
\end{center}
\end{figure}

We perform a quasi-periodic decomposition of the form (\ref{precession}), using the TRIP software (Gastineau and Laskar, 2011), of the JPL/SAT375 ephemeris for the orbit normal motion of Enceladus (see Table \ref{ephem}). This ephemeris is based on actual measurements, so our series expansion may differ from a theoretical secular model. Besides the main precession ($j=1$) characterized by a period of $2.36$ years and an inclination amplitude of $0.00819^\circ$, the series includes four additional terms of smaller amplitudes. \textcolor{black}{The first term in the series expansion is related to the effect of Saturn's oblateness and to the resonance between Enceladus and Dione (Vienne and Duriez, 1995). The third term has a period equal to half the revolution period of Saturn around the Sun and is due to the Sun. The others terms are due to the indirect influence of other satellites of Saturn.} The second term has an amplitude two times smaller than the first term (see Table \ref{ephem}) and contributes significantly to the orbital precession (see Fig. \ref{orbfig}), and therefore potentially to the spin precession.

We also assess if elasticity can trigger a resonant amplification of the solution. For an homogeneous one-layer satellite, the tidal Love number $k_2$ is given by 
\begin{equation}
\label{29} k_2=\frac{3}{2}\frac{1}{1+\frac{19 \mu}{2\bar\rho g R}}\simeq \frac{ 4 \pi G}{19}\frac{\bar\rho^2 R^2 }{\mu},
\end{equation}
with $\mu$ the elastic rigidity and $\bar \rho$ the mean density. Based on Eq. (\ref{29}), we expect that the Love number of a solid Enceladus is more than an order of magnitude smaller than for the large icy satellites as Europa or Ganymede, as a result of Enceladus' smaller size. \textcolor{black}{Using a more realistic two-layers model and rigidities $\mu$ of $3.3$ GPa and $100$ GPa for the ice mantle (e.g. Sotin et al., 1998) and the solid interior (e.g. Turcotte and Schubert, 2002), respectively, we find that $k_2$ ranges from $1.2 \times 10^{-4}$ to $1.9\times 10^{-4}$, depending on the radius of the core}. We will use $k_2=1.5\times 10^{-4}$ in the following, as an estimate for the \textcolor{black}{Love number of a solid Enceladus}. 

If the negative of a forcing frequency $\dot \Omega_j$ is close enough to the free frequency  $\omega_f$ (or $\omega_{f}^{el}$), a resonant amplification of the term $j$ of the rigid (or elastic) solution occurs, since the denominator in Eq. (\ref{rigidfirstorder}) (or Eq. (\ref{elasticfirstorder})) becomes small. However, the \textcolor{black}{free period of a rigid Enceladus ($T_f=0.098$ years) is about one to three orders of magnitude shorter than the negative of the forcing periods $T_{j}$ (see Table 1)}. Therefore, $(\omega_f,\omega_{f}^{el})>>\dot\Omega_j$ and the obliquity amplitudes can be considered as proportional to the free period, at the $10\%$ level at worst. Since the \textcolor{black}{Love number of a solid Enceladus} is very small, the effect of elasticity on the free period is only of $0.01\%$, which is negligible. As a result, no resonant amplification can occur for Enceladus, neither for the rigid case, nor for the elastic case. 

The ratio $C/\tilde C$ differs from unity by about $10^{-6}$, which is even lower than the $0.01\%$ change of the free period. Therefore, as for the \textcolor{black}{free period, the obliquity amplitudes of an elastic Enceladus are about $0.01\%$ larger than their counterparts for a rigid moon. With decreasing ice rigidity, the effect of elastic deformations increases and the obliquity amplitudes become larger. In the limit case where the whole ice mantle deforms as a fluid ($\mu_{ice}\rightarrow 0$ and $k_2\simeq 0.6$), the obliquity amplitudes are about two times larger than the obliquity amplitudes of a rigid moon.  In a more realistic, but already large, range of values for the ice rigidity of $[1,5]$ GPa, $k_2=[1.0,4.6] \times 10^{-4}$ and the obliquity amplitudes are about $[0.007,0.04]\%$ larger than those of a rigid moon.}

\begin{figure}[!htb]
\begin{center}
\includegraphics[width=6cm]{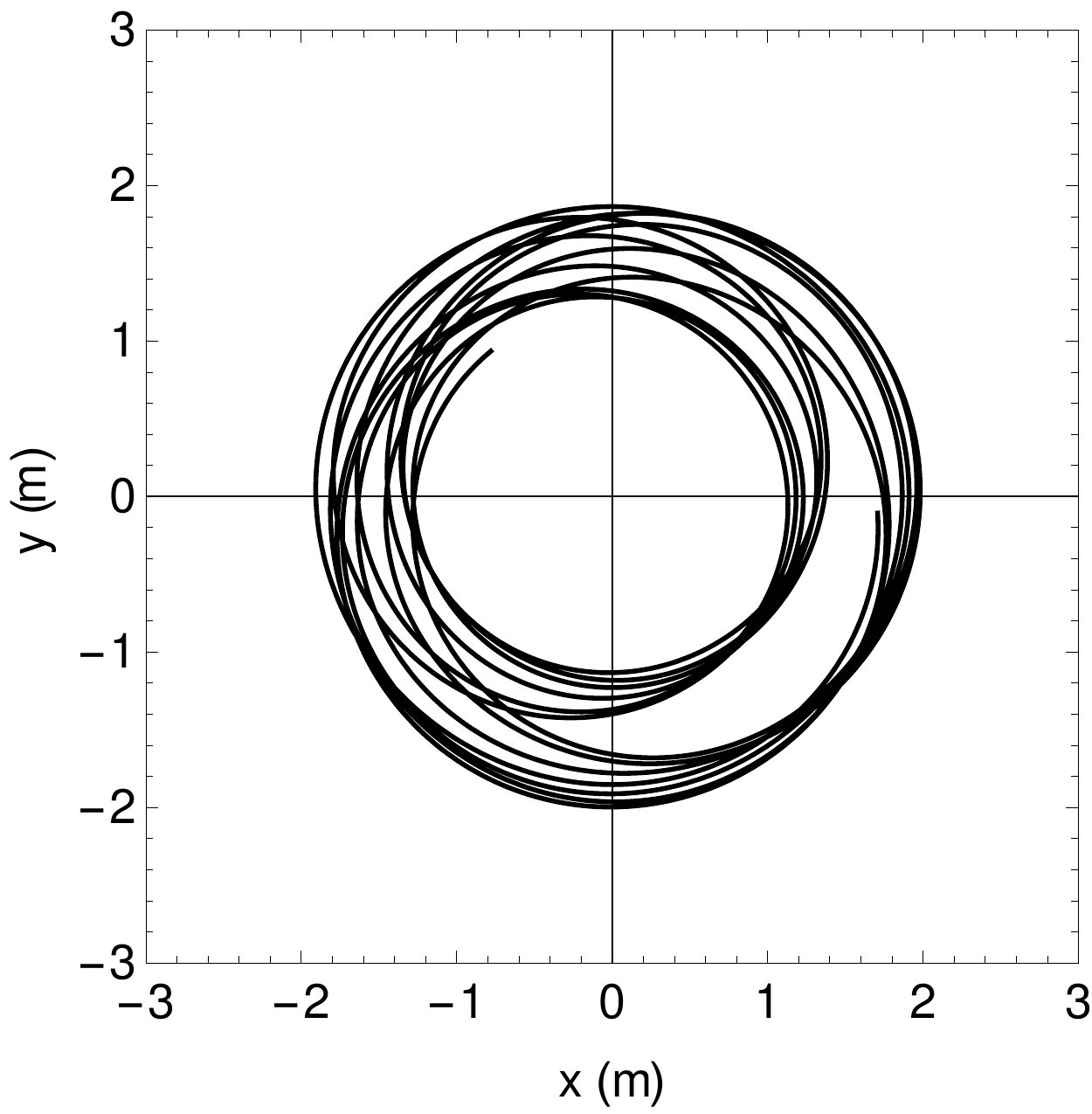}\quad
\includegraphics[width=6cm]{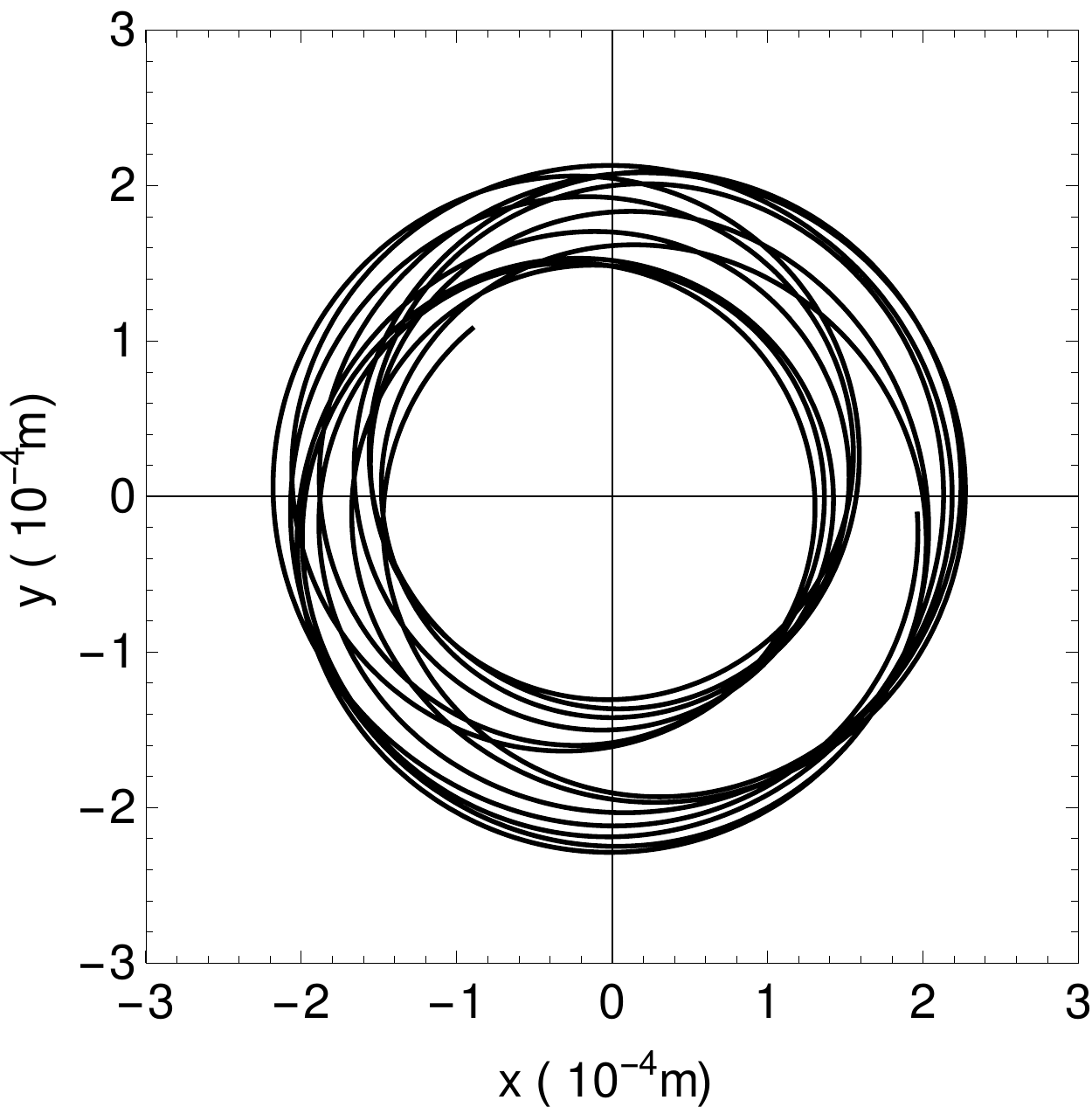}
\caption{\label{spinfig}Difference between the rigid solid spin axis and the orbit axis (left), and between the elastic and rigid solid spin axes (right), projected on the Laplace plane and multiplied by Enceladus's radius $R$. The trajectories are drawn over 30 years, beginning on J2000. The spin precession differs from the orbital precession by about $2$ meters at best, while the effect of elasticity is four orders of magnitude lower. \textcolor{black}{The differences are dominated by the first two frequencies of the series expansions for the orbital and spin precessions.} }
\end{center}
\end{figure}

The obliquity amplitudes $\varepsilon_{j}$ and $\varepsilon^{el}_j$ range from $0.003\times 10^{-4}$ to $3.547\times 10^{-4}$ degrees and are about $25$ to $300$ times smaller than the corresponding inclination amplitudes (see Table \ref{ephem}). \textcolor{black}{Measured at the surface of Enceladus, $\varepsilon_{1}$ and $\varepsilon^{el}_1$ are only about $1.6$ meters, $\varepsilon_{2}$ and $\varepsilon^{el}_2$ are about $0.4$ meters, while the other obliquity amplitudes are even smaller. The first and second terms dominate the solution and the trajectory $S$ of the rigid spin precession projected onto the Laplace plane, computed from Eq. (\ref{forcedsol}), differs from the orbit axis trajectory $N$ by at most about $2$ meters, as can be seen in the left panel of Fig. \ref{spinfig}. The right panel of Fig. \ref{spinfig} demonstrates that the effect of elasticity is four orders of magnitude smaller, since the difference between the trajectories of an elastic and of a rigid moon, computed from  Eqs. (\ref{forcedsol}) and (\ref{forcedelsoleals}), respectively, is at most $0.0002$ meters.} 

\begin{figure}[!htb]
\begin{center}
\includegraphics[width=6cm]{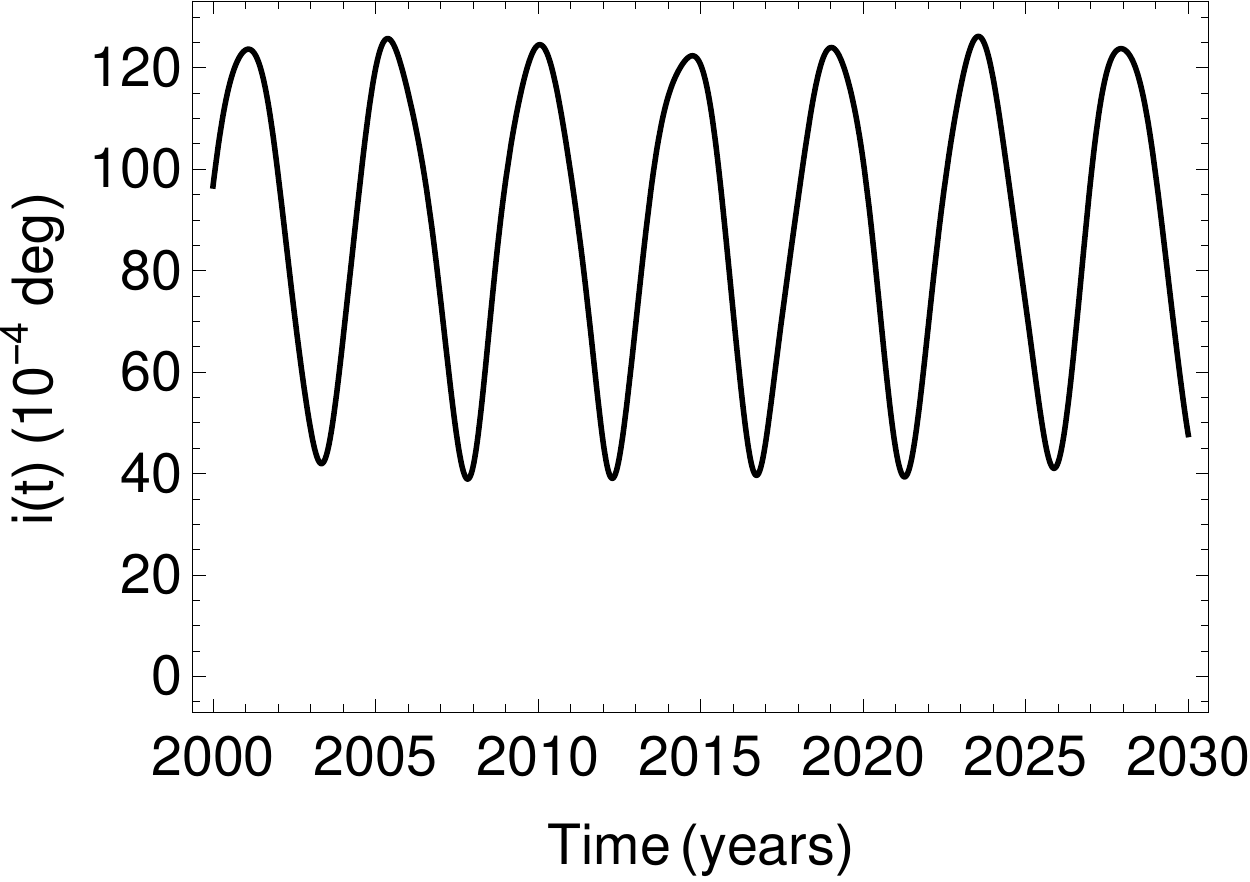}\quad
\includegraphics[width=6cm]{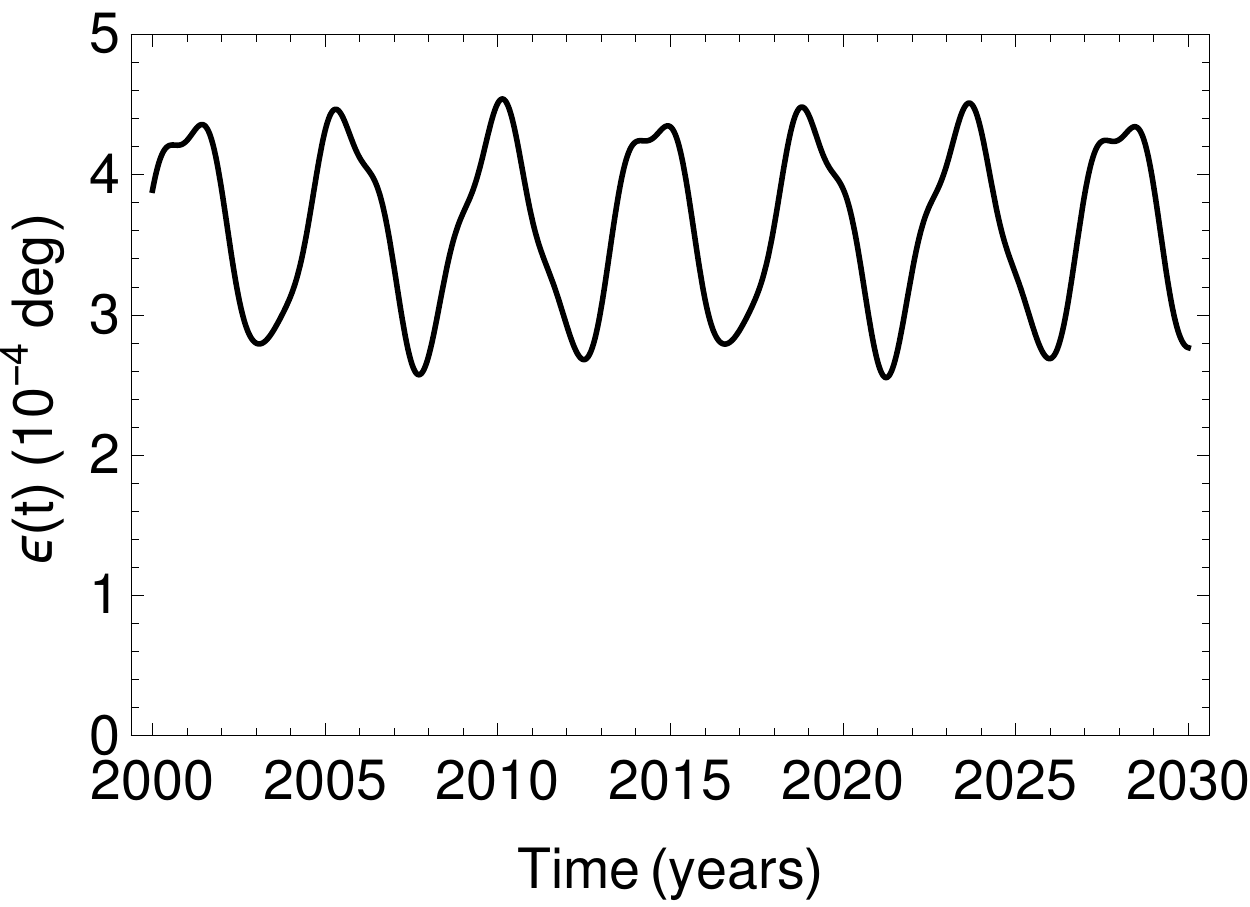}
\caption{\label{inclfig}  \textcolor{black}{Inclination (left) and obliquity of a solid and rigid Enceladus (right) over 30 years beginning on J2000. The obliquity of a solid and elastic Enceladus is not shown here because it would be indistinguishable from the obliquity of the rigid case (only $0.01\%$ of difference).}}
\end{center}
\end{figure}

The time-varying obliquity $\varepsilon(t)$ is about $25$ times smaller than the time-varying inclination $i(t)$ (see Fig. \ref{inclfig}).  $\varepsilon(t)$ varies by about $25\%$ with respect to its mean value and can reach up to $4.48\times 10^{-4}$ degrees. These $25\%$ variations are consistent with the findings of Chen and Nimmo (2011). 

\section{The Cassini state for Enceladus with an internal global ocean and elastic solid layers}
\label{section3}

\subsection{Introduction}

We develop a Cassini state model for Enceladus assuming the satellite is made of three uniform and homogeneous layers of radius $R_j$ and density $\rho_j$: an elastic ice shell (sh), a liquid water ocean (o), and an elastic \textcolor{black}{solid interior (in) that is a mix of ice and silicates}. In principle, the spin orientation of each layer is governed by one angular momentum equation. Saturn exerts an external gravitational torque on each layer, while torques resulting from gravitational and pressure interactions between layers arise. Under the assumption that the liquid ocean is in hydrostatic equilibrium, the pressure torque exerted by the ocean on each adjacent solid layer can be seen as a transfer to the solid layer of the gravitational torques exerted on the part of the ocean aligned with the solid layer, while the total torque on the liquid layer vanishes (Baland et al., 2011). Therefore, the system reduces to the angular momentum equations of the two solid layers. In the rigid case, they can be written, correct up to the first order in spin and orbital inclinations as (see Eqs. (31)-(32) of Baland et al., (2011)):
\begin{eqnarray}
\label{reduce1} C_{sh} \frac{dS_{sh}}{dt}&=&I \kappa'_{sh}(N-S_{sh})-I K (S_{in}-S_{sh}),\\
\label{reduce2} C_{in} \frac{dS_{in}}{dt}&=&I \kappa'_{in}(N-S_{in})+I K(S_{in}-S_{sh}),
\end{eqnarray}
where $S_{sh}$ and $S_{in}$ are the projections on the Laplace plane of the unit vectors along the rotation axes of the shell $\hat s_{sh}$ and of the interior $\hat s_{in}$, respectively. $C_{sh}$ and $C_{in}$ are the polar moment of inertia of the two solid layers. The strength of the internal gravitational torque between the two solid layers is denoted by $K$, while the strengths of the external torques exerted by Saturn on the shell and on the interior are denoted by $\kappa'_{sh}$ and $\kappa'_{in}$, respectively. These internal and external coupling strengths are corrected for the pressure effect (see Eqs. (28)-(30) in Baland et al., 2011). In the following, the torques have to be corrected further for the elastic effect. Besides, as in the solid case, the angular momentum will be affected by the elastic deformations. 

\subsection{External torques}

The external gravitational torque of Saturn on a solid layer is not only exerted on its static bulge, but also on its periodic elastic bulge. As in the case of a solid satellite, only the obliquity component of the periodic bulge of a solid layer plays a role in the angular momentum equations. In a similar way as the forcing strength $\kappa$ for a rigid and solid satellite was changed to $\kappa_{el}$ to account for elasticity (Eq. (\ref{kappael})), the coefficients $\kappa'_{sh}$ and $\kappa'_{in}$ can be easily adapted to the elastic case by considering the \textcolor{black}{contribution $k_2^j$ of each layer $j$ to the total tidal Love number $k_2$,} as done in Van Hoolst et al. (2013) for the study of libration. We then have
\begin{eqnarray}
 \nonumber \kappa_{sh}^{'el}&=&\frac{3}{2}n[(C_{sh}-A_{sh})-k_2^{sh} M_e R^2 q_r\\
 \label{eq21}&&+(C_{o,t}-A_{o,t})-k_2^{o,t} M_e R^2 q_r],\\
 \nonumber \kappa_{in}^{'el}&=&\frac{3}{2}n[(C_{in}-A_{in})-k_2^{in} M_e R^2 q_r\\
 \label{eq22}&&+(C_{o,b}-A_{o,b})-k_2^{o,b} M_e R^2 q_r],
\end{eqnarray}
where $A_{sh}$ and $A_{in}$ are the smallest principal moments of inertia of the two solid layers. The subscripts $(o,t)$ and $(o,b)$ refer to the mass of the ocean aligned with the shell and the interior, respectively, or equivalently, to a top ocean and a bottom ocean above and beneath an arbitrarily chosen sphere inside the ocean.

\subsection{Internal torques}

In the rigid case addressed in Baland et al. (2011), the coupling strength $K$ of the internal gravitational torque, corrected for the pressure effect, between the static bulges of the two solid layers which are misaligned because of their differential spin precession, is defined as
\begin{eqnarray}\label{Kint}
\nonumber K&=&-\frac{8\pi G}{5 n}  M_e R^2 \left[-(C_{20}^{in}+C_{20}^{b,o})[\rho_{sh}(\alpha_{sh}-\alpha_o)+\rho_o \alpha_o]\right.\\
&&\left.+(C_{22}^{in}+C_{22}^{b,o})[\rho_{sh} (\beta_{sh}-\beta_o)+\rho_o\beta_o]\right] ,
\end{eqnarray}
where $\alpha_j$ and $\beta_j$ are the polar and equatorial flattenings of the external interface of layer $j$, defined as the relative differences $((a_j+b_j)/2-c_j)/((a_j+b_j)/2)$ and $(a_j-b_j)/a_j)$ respectively, with $a_j>b_j>c_j$ the radii in the direction of the principal axes of layer $j$. $C_{20}^{j}$ and $C_{22}^{j}$ are the contributions of layer $j$ to the external gravity field coefficients $C_{20}$ and $C_{22}$, which can be expressed in terms of the layer's principal moments of inertia
\begin{equation}
 C_{20}^{j}=-\frac{C_j-\frac{A_j+B_j}{2}}{M_e R^2} \quad \textrm{and} \quad C_{22}^{j}=\frac{B_j-A_j}{4 M_e R^2},
\end{equation}
with $A_j<B_j<C_j$ the principal moments of inertia of layer $j$, defined as (e.g. Van Hoolst et al., 2008)
\begin{eqnarray}
\label{Aj}A_j&=&\frac{8\pi}{15}\rho_j\left[R_j^5\left(1-\frac{1}{3}\alpha_j-\frac{1}{2}\beta_j\right)-R^5_{j-1}\left(1-\frac{1}{3}\alpha_{j-1}-\frac{1}{2}\beta_{j-1}\right)\right],\\
\label{Bj}B_j&=&\frac{8\pi}{15}\rho_j\left[R_j^5\left(1-\frac{1}{3}\alpha_j+\frac{1}{2}\beta_j\right)-R^5_{j-1}\left(1-\frac{1}{3}\alpha_{j-1}+\frac{1}{2}\beta_{j-1}\right)\right],\\
\label{Cj}C_j&=&\frac{8\pi}{15}\rho_j\left[R_j^5\left(1+\frac{2}{3}\alpha_j\right)-R^5_{j-1}\left(1+\frac{2}{3}\alpha_{j-1}\right)\right],
\end{eqnarray}
with the subscript $(l-1)$ referring to the layer located beneath layer $j$.
Note that expression (\ref{Kint}) has been corrected with respect to the expression used in Baland et al. (2011, 2012, 2014). \textcolor{black}{This correction has limited impact on the results reported in these studies.} The reader is referred to Appendix B for a formal demonstration.

A similar coupling coefficient for the torque between static bulges misaligned because of the differential rotation rate of the solid layers has been defined in libration studies (e.g. Van Hoolst et al., 2009). The existence of periodic elastic bulges led us to consider also internal torques between static and periodic bulges, \textcolor{black}{while the torque between the two periodic bulges has been neglected}. The sum of these additional torques exerted on the interior $\delta \Gamma$ is given by Eq. (\ref{C30}), and can be projected on the Laplace plane as
\begin{equation}
 \label{Eq37}\delta \Gamma=I K_{pi} (S_{in}-N)+I K_{ps} (S_{sh}-N)
\end{equation}
The additional torque on the shell is equal and opposite to the additional torque on the interior. Additional coupling strengths were defined to account for the torque between the static bulge of the shell/interior and the librational tidal bulge of the interior/shell (Van Hoolst et al., 2013). Here, the periodic tidal bulges induced by obliquity tides lead to the consideration of a coupling strength for the torque between the periodic bulge of the interior and the static bulge of the shell ($K_{pi}$) and of another coupling strength for the torque between the periodic bulge of the shell and the static bulge of the interior ($K_{ps}$), which are expressed as (see Appendix C for the demonstration)
\begin{eqnarray}
\label{Kpi} K_{pi}&=&\frac{3}{2 n}(k_2^{in}+k_2^{o,b}) M_e R^2 q_r (\phi_{20}^{sh}-2\phi_{22}^{sh}+\phi_{20}^{o,t}-2\phi_{22}^{o,t}),\\
\nonumber K_{ps}&=&\frac{6\pi G}{5}n R^2  \left[\rho_{sh}(\frac{y_{sh}}{R_{sh}} -\frac{y_o}{R_o})+\rho_o \frac{y_o}{R_o} \right]\\
\label{Kps} &&[(C_{in}-A_{in})+(C_{o,b}-A_{o,b})],
\end{eqnarray}
where $\phi_{20}^{j}$ and $\phi_{22}^{j}$ are the contribution of layer $j$ to the coefficients of the internal gravity field that can be expressed in terms of the flattenings of the layer's interfaces (see Eqs. (\ref{phi20}) and (\ref{phi22})) and where $y_j$ is the radial displacement at the top interface of layer $j$, calculated for a tidal potential with degree-two component equal to $-1$ m$^2$/s$^2$ at the surface. 
  
\subsection{Angular momentum equations}

By using the expressions for the coupling strengths  derived in the previous subsections, and by correcting the angular momentum of the shell and of the interior in the same way as in the solid case (see Eq. (\ref{26})), the angular momentum equations (\ref{reduce1}) and (\ref{reduce2}) can be corrected for elasticity and become
\begin{eqnarray}
\nonumber \tilde C_{sh}\frac{dS_{sh}}{dt}&=&I \kappa_{sh}^{'el} (N-S_{sh})-I K (S_{in}-S_{sh})-I K_{pi} (S_{in}-N)\\
\label{30}&&-I K_{ps} (S_{sh}-N)+(\tilde C_{sh}-C_{sh})\frac{d N}{dt},\\
\nonumber \tilde C_{in}\frac{dS_{in}}{dt}&=&I \kappa_{in}^{'el} (N-S_{in})+I K (S_{in}-S_{sh})+I K_{pi} (S_{in}-N)\\
\label{31}&&+I K_{ps} (S_{sh}-N)+(\tilde C_{in}-C_{in})\frac{d N}{dt},\\
\label{43}\tilde C_{sh} &=&\, C_{sh} -\frac{1}{2}k_2^{sh} M_e R^2 q_r,\\
\label{44}\tilde C_{in} &=&\, C_{in} -\frac{1}{2}k_2^{in} M_e R^2 q_r.
\end{eqnarray} 
In the right hand members of Eqs. (\ref{30})-(\ref{31}), the first term, proportional to $\kappa_{sh}^{'el}$ or $\kappa_{in}^{'el}$, is the external gravitational torque, corrected for the pressure effect, exerted by Saturn on the elastic shell or the elastic interior, respectively. The second term, proportional to $K$, is the internal gravitational torque between the static bulges of the solid layers, corrected for the pressure effect. The third and fourth terms, proportional to $K_{pi}$ and $K_{ps}$, are the additional torques between the static and periodic bulges, also corrected for the pressure effect. The last terms come from the change in angular momentum due to obliquity tides. The system can be rearranged as:
\begin{eqnarray}
\label{arranged1}\tilde C_{sh}\frac{dS_{sh}}{dt}&=&I \tilde\kappa_{sh} (N-S_{sh})-I \tilde K_{sh} (S_{in}-S_{sh})+(\tilde C_{sh}-C_{sh})\frac{d N}{dt},\\
\label{arranged2}\tilde C_{in}\frac{dS_{in}}{dt}&=&I \tilde\kappa_{in} (N-S_{in})+I \tilde K_{in} (S_{in}-S_{sh})+(\tilde C_{in}-C_{in})\frac{d N}{dt},
\end{eqnarray} 
with 
\begin{eqnarray}
 \tilde\kappa_{sh}&=&\kappa_{sh}^{'el}+ K_{ps}+K_{pi},\\
 \tilde\kappa_{in}&=&\kappa_{in}^{'el}-K_{ps}- K_{pi},\\
 \tilde K_{sh}&=&K+K_{pi},\\
 \tilde K_{in}&=&K-K_{ps}.
\end{eqnarray}

\subsection{Free modes and forced solution}

The system (\ref{arranged1}-\ref{arranged2}) has two free modes which can be found by setting $N$ equal to $(0,0)$ and which are characterized by two free frequencies $\omega_{\pm}^{el}$ defined as
\begin{eqnarray}
\label{freemodes}\omega_{\pm}^{el}&=&-(Z \pm \sqrt{\Delta})/(2 \tilde C_{in} \tilde C_{sh}),\\
Z&=&(\tilde C_{in}\tilde K_{sh} +\tilde C_{sh}\tilde K_{in})-\tilde C_{sh} \tilde\kappa_{in} -\tilde C_{in}\tilde \kappa_{sh}, \\
\label{freemodes3}\Delta&=&-4 \tilde C_{in} \tilde C_{sh} (- \tilde K_{sh} \tilde\kappa_{in}-\tilde K_{in} \tilde\kappa_{sh}+\tilde\kappa_{in} \tilde\kappa_{sh} )+Z^2. 
\end{eqnarray}
The free frequency $\omega_+^{el}$ corresponds to a mode where both the solid layers precess in phase with each other, while they are out of phase for $\omega_-^{el}$. The \textcolor{black}{free frequencies of a satellite with rigid solid layers $\omega_{\pm}$} (Eq. (33) of Baland et al., 2011) can be retrieved by setting $k_2^j$ and $y_j$ to zero in the coefficients of Eqs. (\ref{freemodes})-(\ref{freemodes3}). 

\textcolor{black}{For the orbital precession given by Eq. (\ref{precession}) and by injecting forced solutions for the spin precessions of the form
\begin{eqnarray}
S_{sh}&=&\sum_j (i_j+\varepsilon_{j,sh}^{el}) e^{I (\dot \Omega_j t+ \gamma_j-\pi/2)},\\
S_{in}&=&\sum_j (i_j+\varepsilon_{j,in}^{el}) e^{I (\dot \Omega_j t+ \gamma_j-\pi/2)},
\end{eqnarray}
in Eqs. (\ref{arranged1})-(\ref{arranged2}),} we have, correct up to the first order in inclination and obliquity amplitudes, the following shell obliquity amplitudes $\varepsilon_{j,sh}^{el}$ and interior obliquity amplitude $\varepsilon_{j,in}^{el}$ 
\begin{eqnarray}
\label{solution}
\varepsilon_{j,sh}^{el}&=&\frac{i_j \dot\Omega_j (C_{in}\tilde K_{sh} +C_{sh}\tilde K_{in}-C_{sh} \tilde \kappa_{in}-\tilde C_{in}\, C_{sh} \,\dot\Omega_j)}{\tilde C_{in}\tilde C_{sh}(\omega_{+}+\dot\Omega_j)(\omega_{-}+\dot\Omega_j)},\\
\label{solution2}\varepsilon_{j,in}^{el}&=&\frac{i_j \dot\Omega_j (C_{in}\tilde K_{sh} +C_{sh}\tilde K_{in}-C_{in} \tilde \kappa_{sh}-C_{in}\, \tilde C_{sh} \,\dot\Omega_j)}{\tilde C_{in}\tilde C_{sh}(\omega_{+}+\dot\Omega_j)(\omega_{-}+\dot\Omega_j)}.
\end{eqnarray}
The rigid obliquity amplitudes $\varepsilon_{j,sh}$ and $\varepsilon_{j,in}$ (Eqs. (36)-(37) of Baland et al., 2012) can be retrieved by setting $k_2^j$ and $y_j$ to zero \textcolor{black}{in the expressions for the different coefficients of} Eqs (\ref{solution}-\ref{solution2}). 

\textcolor{black}{The actual time-varying obliquity of the elastic shell and interior, $\varepsilon^{el}_{sh}(t)$ and $\varepsilon^{el}_{in}(t)$, are defined similarly as in Eq. (\ref{incl}): 
\begin{equation}\label{actualobl}
\varepsilon^{el}_{sh}(t)\simeq\|S_{sh}-N\|\quad\textrm{and}\quad \varepsilon^{el}_{in}(t)\simeq\|S_{in}-N\|,
\end{equation}}

\subsection{Results}

\subsubsection{Density profiles and static flattenings}
\label{profiles}

\begin{table}[htdp]
\begin{center}
\begin{tabular}{lll}
\hline
layer & Thickness[km]/Radius[km] & Density[kg\, m$^{-3}$] \\
\hline
ice shell		& $h_{sh}=\{5, 6, ..., 50\}$ 	& $\rho_{sh}=\{900, 925, ..., 1000\}$  \\
ocean			& $h_o=2.1-67.1$		& $\rho_{o}=\{\textcolor{black}{950, 975} ..., 1050\}$ \\
solid interior		& $R_{in}=\{180,185, ..., 200\}$ 		& $\rho_{in}=2246.1-2602.9$ \\
\hline
\end{tabular}
\end{center}
\caption{\label{tableEnceladus}The ranges of size and of density of the three internal layers of Enceladus, consistent with the given constraints on mass, radius, and mean moment of inertia. Within a given set denoted by curly brackets, the values are equally spaced. \textcolor{black}{This table defines 2191 solutions for the density profiles.}} 
\end{table}

We build a large set of density profiles for Enceladus (see Table \ref{tableEnceladus}) \textcolor{black}{by extending the two-layers density profile proposed by Iess et al. (2014). In their model, the H$_2$O mantle has a uniform density of 1000 kg m$^{-3}$, while the core has density and radius of about $2400$ kg m$^{-3}$ and $190$ km, respectively. In our three-layers density profiles, the water layer is divided into a solid and a liquid part that can have different densities. For pure water, and neglecting the small effects of pressure and temperature, the density of the shell $\rho_{sh}$ and of the ocean $\rho_o$ would be $920$ kg m$^{-3}$ and $1000$ kg m$^{-3}$, respectively. Here we consider a range for $\rho_{sh}$ from $900$ to $1000$ kg m$^{-3}$, and a range for $\rho_o$ from $950$ to $1050$ kg m$^{-3}$, to take into account effects of possible impurities (ammonia and/or salts) and porosity}. We consider values for the ice shell thickness $h_{sh}$ equally spaced by $1$ km, between $5$ and $50$ km. The values for the interior radius $R_{in}$ are between $180$ and $200$ km, and are equally spaced by $5$ km, \textcolor{black}{which allows some variations with respect to the density profile proposed by Iess et al. (2014).} The density of the interior $\rho_{in}$ and the thickness of the ocean $h_o$ are computed from the constraints on the mass $M_e$ and on the radius $R$ and from the values of the other parameters. Only density profiles with a mean moment of inertia $I/M_eR^2$ compatible with the estimate $I/M_eR^2=0.335\pm0.005$ (2$\sigma$) of Iess et al. (2014) are considered. The density of the interior $\rho_{in}$ is then between $2246$ kg m$^{-3}$ and $2603$ kg m$^{-3}$, \textcolor{black}{a range that includes the value of Iess et al. (2014)}. This procedure provides \textcolor{black}{2191} solutions for the density profiles. 

In this section, we assume that Enceladus does not deviate from hydrostatic equilibrium. Following Van Hoolst et al. (2008), the static triaxial ellipsoidal shape of any equipotential surface of constant density can be expressed as 
\begin{equation}
r=r_0\left(1-\frac{2}{3}\alpha(r_0) P_2^0(\cos\varphi)+\frac{1}{6}\beta(r_0) P_2^2(\cos\varphi)\cos2\lambda\right),
\end{equation}
with $r$, $\varphi$ and $\lambda$ the spherical coordinates (radius, colatitude and longitude) of a point on the deformed equipotential surface, $r_0$ the mean radius, and $\alpha(r_0)$ and $\beta(r_0)$ the polar and equatorial flattenings of the ellipsoidal surface. For a synchronous satellite, $\beta(r_0)=\frac{6}{5}\alpha(r_0)$. $P_2^0(\cos\varphi)$ and $P_2^2(\cos\varphi)$ are associated Legendre functions. The polar flattenings $\alpha_j$ (and so the equatorial flattenings $\beta_j=\frac{6}{5}\alpha_j$) of the different layers, needed to compute the coupling strengths $\kappa^{'el}_{sh}, \kappa^{'el}_{in}$ of Eqs. (\ref{eq21})-(\ref{eq22}) and $K$ of Eq. (\ref{Kint}), are obtained by solving Clairaut's equation 
\begin{equation}
 \frac{d^2\alpha}{dr_0^2}+\frac{6}{r_0}\frac{\rho}{\bar\rho}\frac{d\alpha}{dr_0}-\frac{6}{r_0^2}\left(1-\frac{\rho}{\bar\rho}\right)\alpha=0,
\end{equation}
with the following boundary condition
\begin{equation}
 \frac{d\alpha}{dr_0}(R)=\frac{1}{R}\left[\frac{25}{4}q_r-2\alpha(R)\right]. 
\end{equation}
$\bar\rho$ is the mean density inside the sphere of radius $r_0$. In Clairaut's theory, the hydrostatic polar flattening of the surface is defined as
\begin{equation}
 \alpha_{sh}=\alpha(R)=-\frac{3}{2}C^{HE}_{20}+\frac{5}{4}q_r,
\end{equation}
with $C^{HE}_{20}=-\frac{10}{3}C^{HE}_{22}$. \textcolor{black}{The expected hydrostatic coefficients $C^{HE}_{20}$ and $C^{HE}_{22}$ are related to the mean moment of inertia $I/M_eR^2$ of the chosen density profile through Radau's equation, which is given by} 
\begin{equation}
 \frac{I}{M_e R^2}=\frac{2}{3}\left[1-\frac{2}{5}\left(\frac{25}{4}\frac{q_r}{-\frac{3}{2}C^{HE}_{20}+\frac{5}{4}q_r}-1\right)^{1/2}\right].
\end{equation}
$C^{HE}_{20}$ and $C^{HE}_{22}$ may differ from the actual measured gravity coefficients, since Enceladus deviates from hydrostatic equilibrium, as demonstrated by Iess et al. (2014) and McKinnon (2015). The effect of this departure on the upper bound of Enceladus' obliquity will be discussed in section 4. 

\subsubsection{Elastic deformations} 

The radial tidal displacements $y_j$ at the outer surface of the layers $j$ are calculated by solving the equation of motion, Poisson's equation, and the continuity equation for a spherically symmetric satellite. \textcolor{black}{In addition, we have used the constitutive equation between stress and strain for an isotropic elastic solid.} The set of differential equations solved for tides is formally exactly the same as for free seismic oscillations (see, e.g., Dahlen and Tromp 1999) by replacing the Eulerian perturbation of the gravitational potential by the sum of that perturbation and the tidal potential. We calculated the $y_j$ for a tidal potential with degree-two components equal to $-1$~m$^2$/s$^2$ at the surface. The only difference with respect to seismic normal modes is that the surface boundary condition on the \textcolor{black}{radial derivative of the} gravitational potential becomes non-homogeneous (e.g. Tobie et al., 2005; Rivoldini et al., 2009). As the libration period is large compared to the seismic mode frequencies and the effect of inertial forces is small, the Love numbers are almost independent of frequency and we calculate the Love numbers for zero frequency. \textcolor{black}{We do not include viscoelasticity as its effect on the amplitude of the Love numbers for realistic viscosity parameters is small and well below the effect of uncertainties in model parameters, such as the ice rigidity.}  We also assume incompressibility in our calculation as did, e.g., Moore and Schubert (2000) and Wahr et al. (2006). 

\textcolor{black}{The contribution $k_2^j$ of each layer $j$ to the total tidal Love number $k_2$} can be expressed as functions of the radial tidal displacement of the interfaces $y_j$:
\begin{eqnarray}
 \label{k2sh}k_2^{sh}&=&\frac{4\pi G\rho_{sh}}{5R^3}\left(R^4 y_{sh}-R_o^4y_o\right),\\
 \label{k2ot}k_2^{o,t}&=&\frac{4\pi G\rho_{o}}{5R^3}R_o^4 y_{o},\\
 \label{k2ob}k_2^{o,b}&=&-\frac{4\pi G\rho_{o}}{5R^3}R^4_{in}y_{in},\\
 \label{k2in}k_2^{in}&=&\frac{4\pi G\rho_{in}}{5R^3}R^4_{in} y_{in}.
\end{eqnarray}
$y_j$ and $k_2^j$ are used to compute the torque strengths $\kappa^{'el}_{sh}, \kappa^{'el}_{in}$ of Eqs. (\ref{eq21})-(\ref{eq22}) and $K_{pi}$ and $K_{ps}$ of Eqs (\ref{Kpi}-\ref{Kps}), as well as the effect of deformations on the effective polar moment of inertia of the shell $\tilde C_{sh}$ and of the interior $\tilde C_{in}$ (Eqs. (\ref{43})-(\ref{44})).

\begin{figure}[!htb]
\begin{center}
\includegraphics[height=5cm]{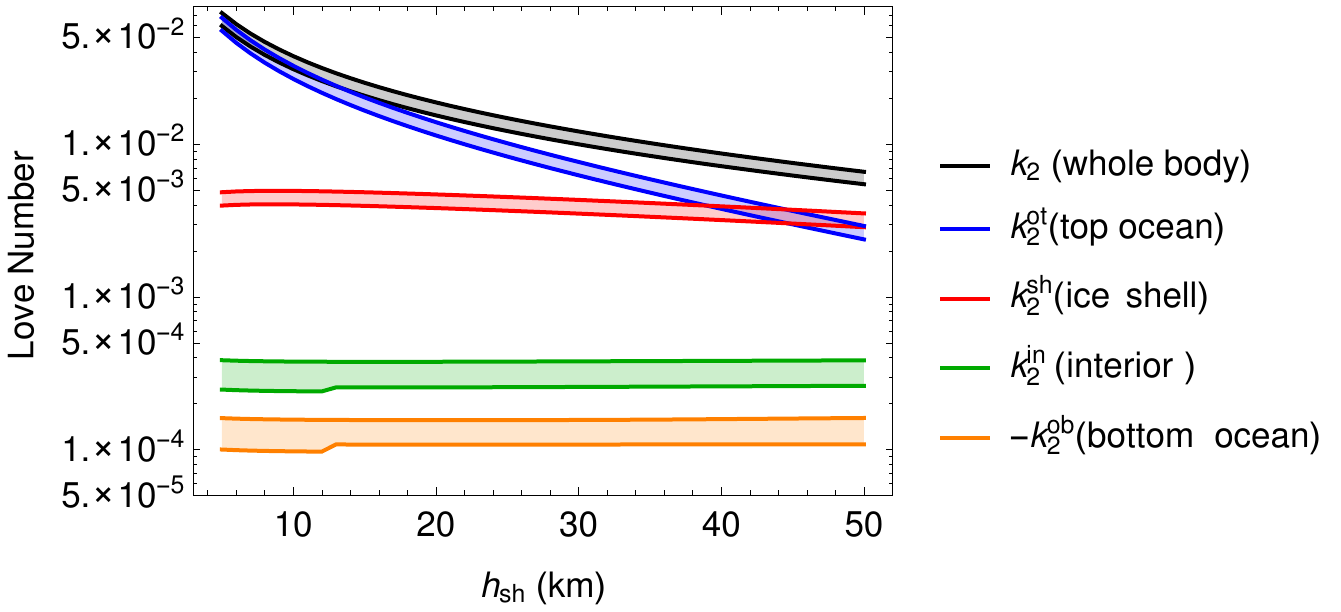}
\caption{\label{k2layers} \textcolor{black}{Ranges of the Love numbers $k_2$ of the whole Enceladus, and of the contributions of the top part of the ocean $k_2^{o,t}$, of the ice shell $k_2^{sh}$, of the interior $k_2^{in}$, and of the bottom part of the ocean $k_2^{o,b}$ as a function of shell thickness $h_{sh}$ for all the density profiles defined in Table \ref{tableEnceladus}, and for $\mu_{sh}=3.3$ GPa and $\mu_{in}=100$ GPa}. $k_2^{in}$ and $k_2^{o,b}$ are very small and of different signs. }
\end{center}
\end{figure}

\textcolor{black}{For the density profiles of Enceladus considered in Table \ref{tableEnceladus} and for rigidities $\mu_{sh}=3.3$ GPa and $\mu_{in}=100$ GPa, the total Love number $k_2$ ranges from $0.005$ for thick ice shells to $0.072$ for thin ice shells (see black region in Fig. \ref{k2layers}), and is $30$ to $500$ times larger than the chosen value for a solid Enceladus ($1.5\times 10^{-4}$). For $\mu_{sh}=1$ GPa ($5$ GPa), $k_2$ ranges from $0.017$ ($0.004$) for thick ice shells to $0.072$ ($0.050$) for thin ice shells. $k_2$ depends mainly on the ice shell thickness $h_{sh}$, and weakly on other parameters like the shell and ocean densities. The shell (red region in Fig. \ref{k2layers}) and the top ocean (blue region) contribute together at least $95.6\%$ to the total Love number and up to $99.8\%$, since the solid shell top and bottom interfaces are more deformed by the tides than the interior. The thinner the shell is, the larger $y_{sh}$ and $y_o$ are and the closer they are to each other. Therefore, the contribution of the top ocean to the total Love number is maximal when the shell is thin (see Eqs. (\ref{k2sh})-(\ref{k2ot})). For $h_{sh}=5$ km, the top ocean contribution is about $93\%$ while the shell contribution is about $6.5\%$. The respective contributions are about $44\%$ and $52.5\%$ for $h_{sh}=50$ km. The joint contributions of the interior (green region) and of the bottom ocean (orange region) is at best of $4.4\%$ when the shell is thick.}
 
\subsubsection{Solution \textcolor{black}{for the spin precession}}
\label{computation}

\begin{figure}[!htb]
\begin{center}
\includegraphics[height=5.5cm]{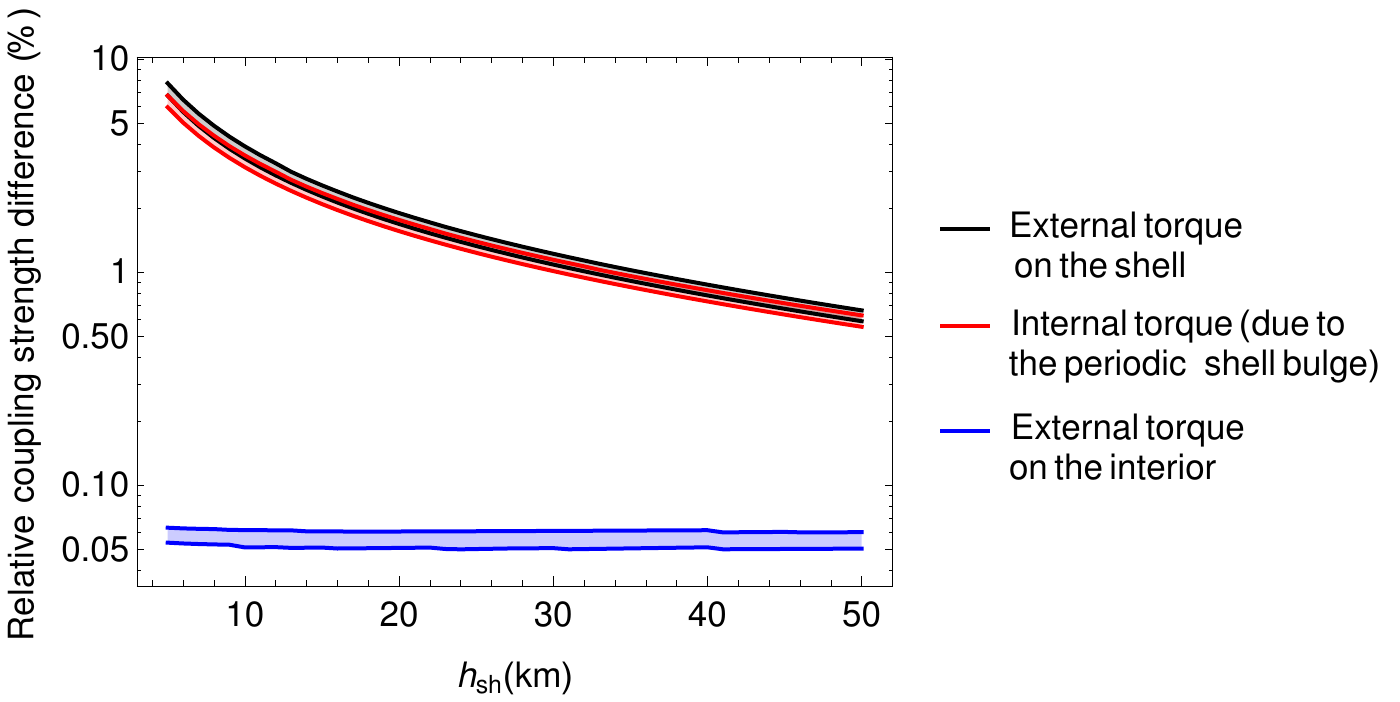}
\caption{\label{kapparel} \textcolor{black}{Ranges of the effect of elastic deformations on the coupling strengths, for the density profiles defined in Table \ref{tableEnceladus}. Black and blue: relative influence of elastic deformations on the strengths of the external torques, corrected for the pressure, on the shell and on the interior ($|(\kappa^{'el}_{sh}-\kappa'_{sh})/\kappa^{'el}_{sh}|$ and $|(\kappa^{'el}_{in}-\kappa'_{in})/\kappa^{'el}_{in}|$), as a function of the shell thickness. Red: ratio of the strength of the additional internal gravitational torque involving the periodic bulge of the shell over the strength of the internal gravitational torque between the static bulges of the solid layers ($|K_{ps}/K|$). The ratio $|K_{pi}/K|$ is of the order of $10^{-10}\%$, and hence not shown here.}}
\end{center}
\end{figure}  
 
The effect of elastic deformations on the solutions (\ref{solution}-\ref{solution2}) can be anticipated by an assessment of their effect on the coupling strengths. The relative difference between the rigid and elastic versions of the strengths of the external torques, corrected for the pressure, is between $0.6\%$ and $7.7\%$ for the shell ($|(\kappa^{'el}_{sh}-\kappa'_{sh})/\kappa^{'el}_{sh}|$) and lower ($\sim 0.06\%$) for the interior ($|(\kappa^{'el}_{in}-\kappa'_{in})/\kappa^{'el}_{in}|$) since the deformations of the interior are small \textcolor{black}{(see black and blue regions in Fig. \ref{kapparel})}. The effect of elastic deformations on the external torques is then larger than in the solid case, for which it was of about $0.01\%$. We also compare the additional elastic internal coupling strengths $K_{pi}$ and $K_{ps}$ to the rigid internal coupling strength $K$. $|K_{ps}/K|$ decreases with increasing ice shell thicknesses from $0.5\%$ to $6.7\%$ \textcolor{black}{(see red region in Fig. \ref{kapparel})}, while $K_{pi}$ is completely negligible in front of $K$ (\textcolor{black}{of the order $10^{-10}\%$, and hence not shown in Fig. 5.}), meaning that the additional internal gravitational torque  involving the periodic bulge of the interior (first term of Eq. (\ref{Eq37})) is quasi negligible. Contrary to the solid case, we then expect a small but non-negligible effect of elastic deformations on the solution for the spin precession, especially for thin ice shells, thanks to the elastic deformations of the shell top and bottom interfaces.

\begin{figure}[!htb]
\begin{center}
\includegraphics[height=5cm]{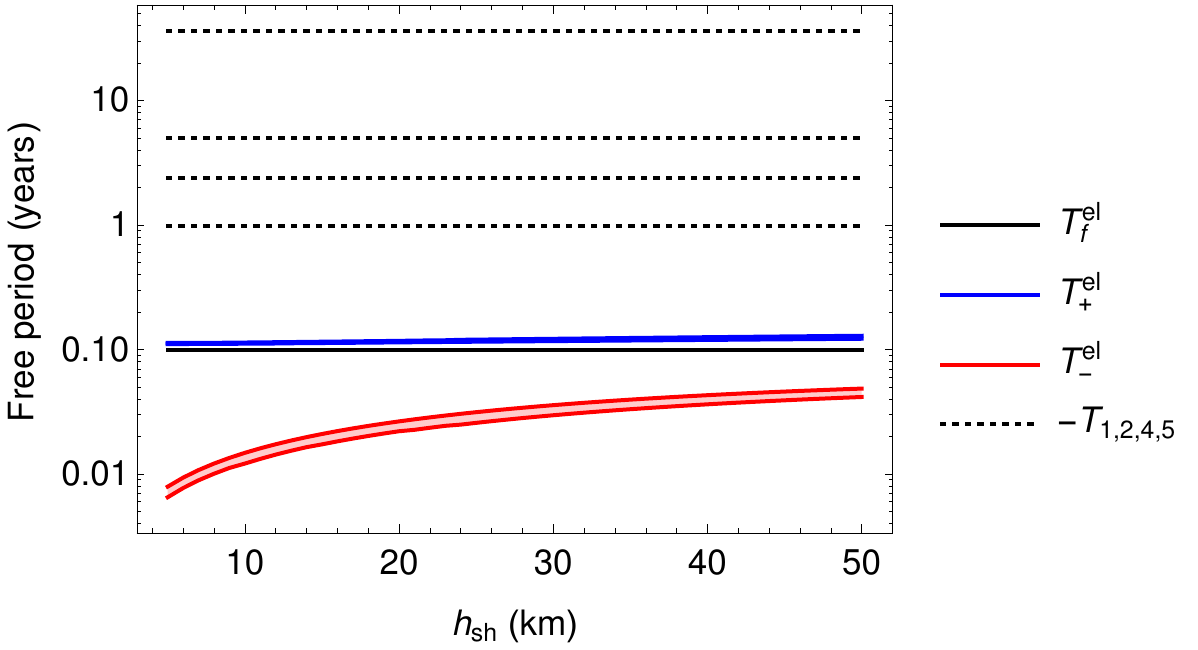}
\caption{\label{tptm}\textcolor{black}{Ranges of the free periods $T_{\pm}^{el}$ as a function of the shell thickness $h_{sh}$ for the density profiles defined in Table \ref{tableEnceladus} (blue and red regions), compared to the free period $T^{el}_f$ of the solid case (solid black line) and the opposite of the negative forcing periods $T_{j=1,2,4,5}$  (dotted black lines).}}
\end{center}
\end{figure}
% \newpage

\textcolor{black}{As in the solid case, we now investigate whether a resonant amplification of one term of the solution for the shell spin precession is possible. The free period $T_+^{el}=2\pi/\omega_{+}^{el}$ is of the order of a tenth of a year, and is, as expected, close to the free period of the solid case of $0.098$ years (see blue region in Fig. \ref{tptm}), while the free period $T_-^{el}=2\pi/\omega_{-}^{el}$ is about two to ten times shorter than $T_+^{el}$ (red region). Both free periods $T_{\pm}^{el}$ increase with increasing ice shell thicknesses $h_{sh}$. However, neither the presence of the ocean, nor elasticity, leads to a resonant amplification, since the opposite of the forcing periods $T_{j=1,2,4,5}$ are significantly larger than the free periods (see Fig. \ref{tptm})}.

\begin{figure}[htb]
\begin{center}
\quad \includegraphics[height=3.9cm]{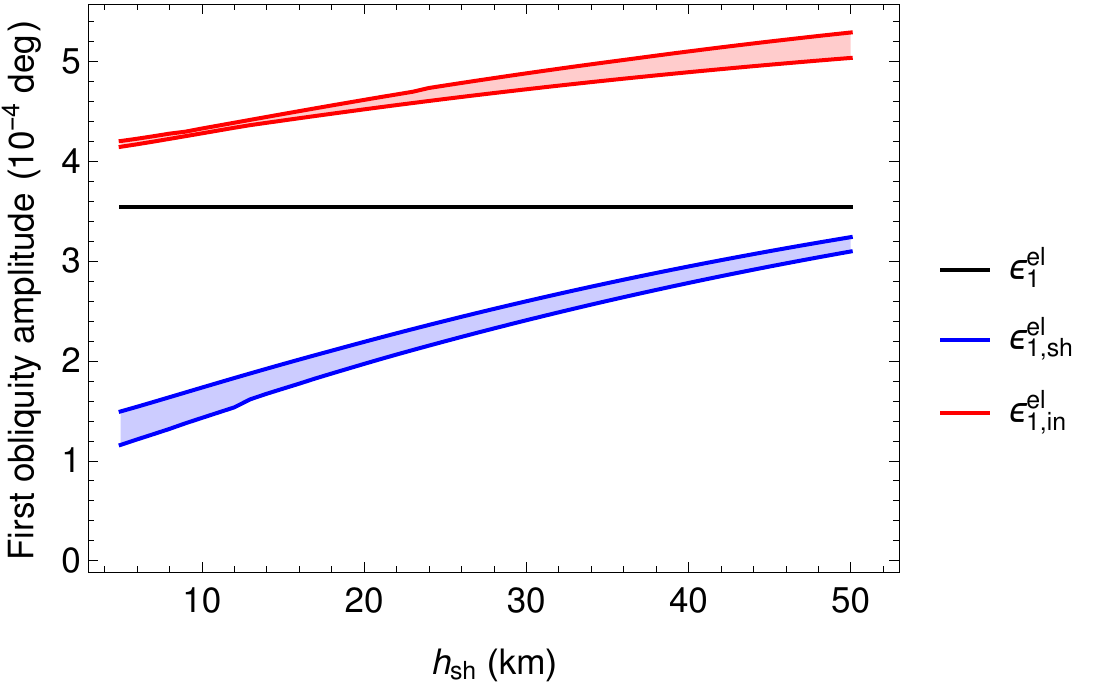}\quad 
\includegraphics[height=4cm]{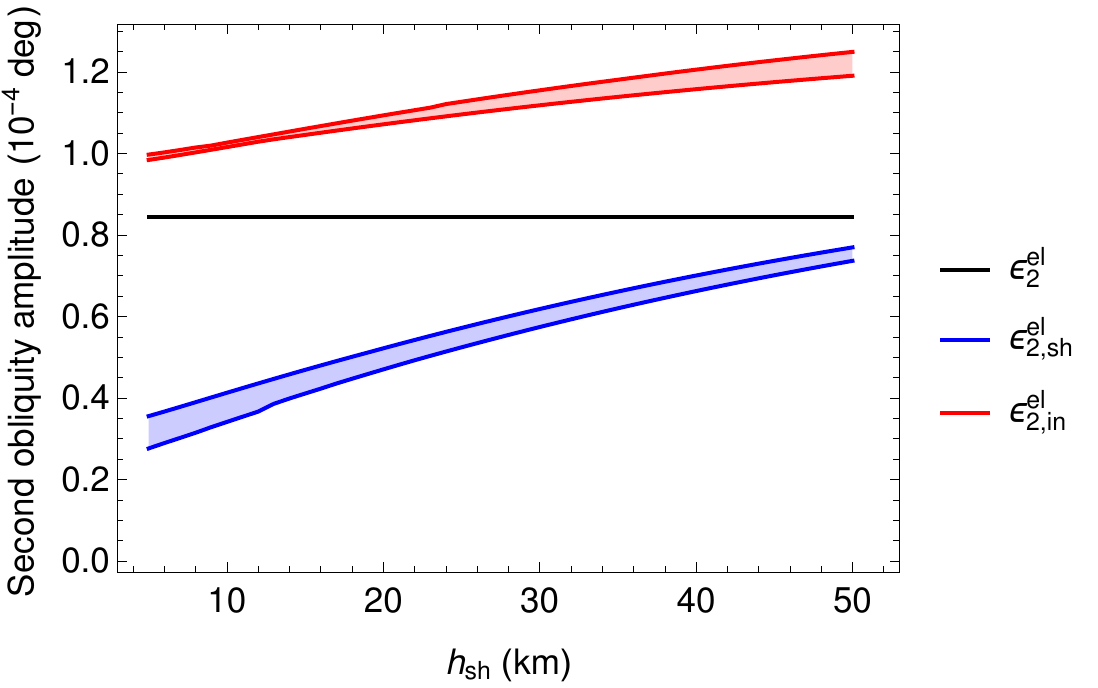}
\includegraphics[height=4cm]{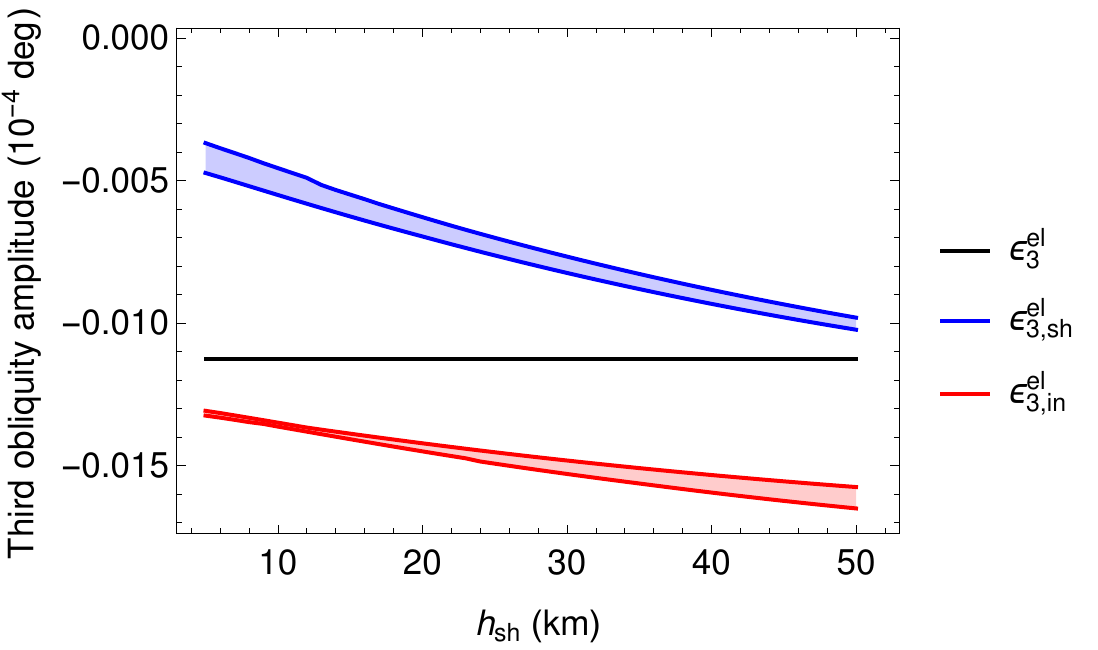}
\includegraphics[height=3.95cm]{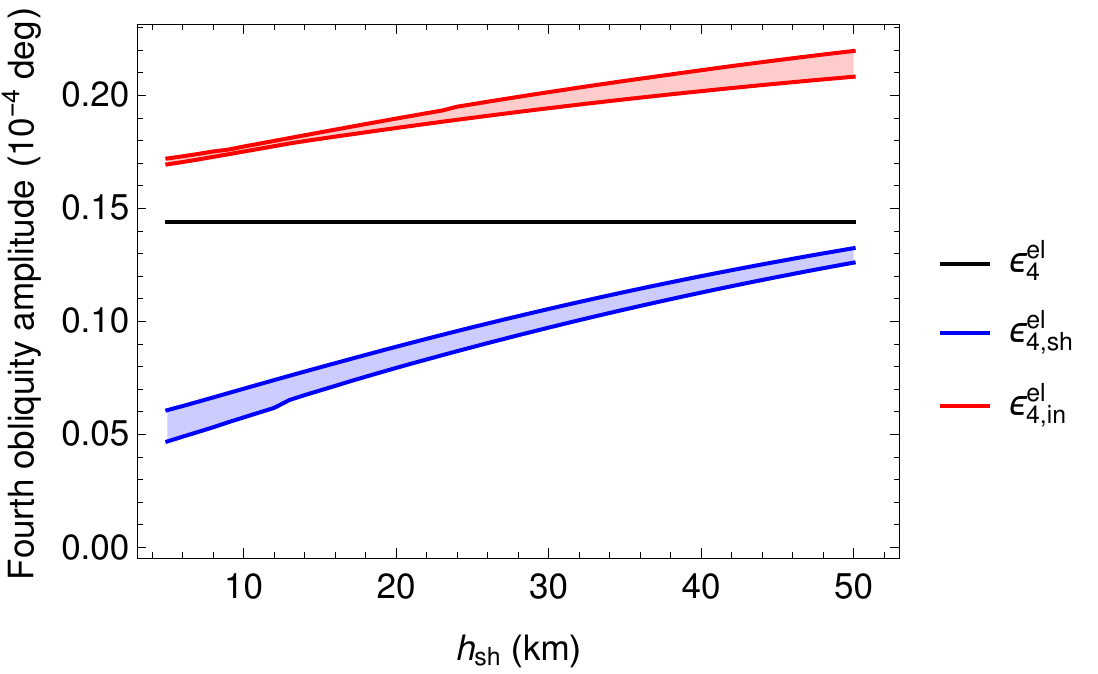}
\includegraphics[height=3.9cm]{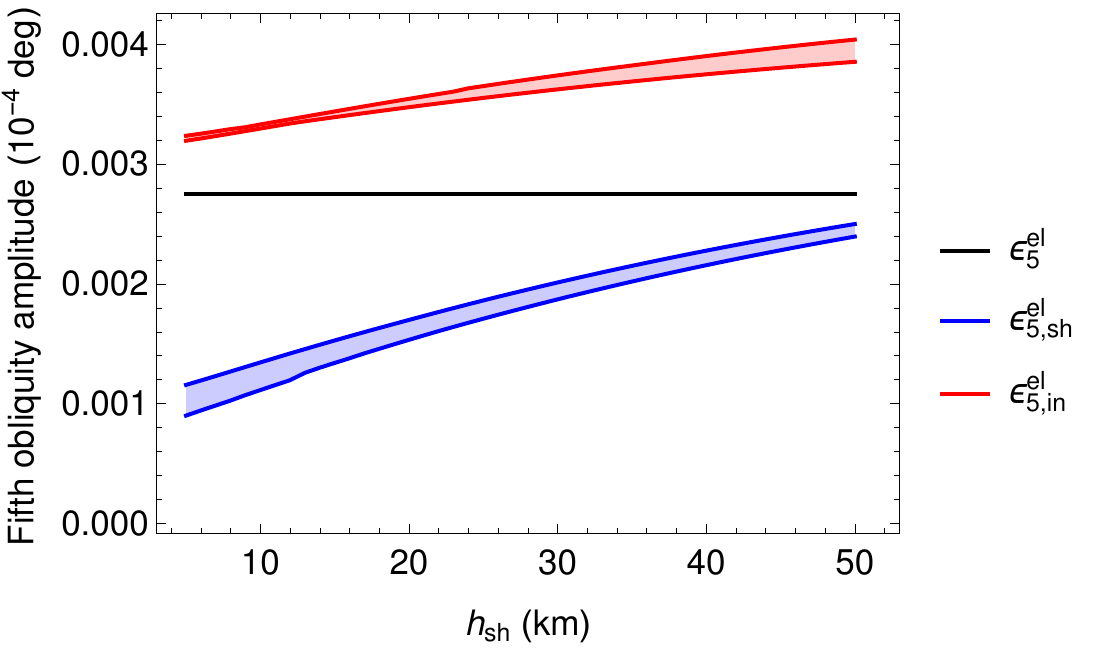}
\caption{\label{esei} \textcolor{black}{Ranges of the obliquity amplitudes of the elastic shell ($\varepsilon_{j,sh}^{el}$) and of the interior ($\varepsilon_{j,in}^{el}$), as a function of shell thickness $h_{sh}$ for the density profiles defined in Table \ref{tableEnceladus}, compared to obliquity amplitudes of en entirely solid Enceladus ($\varepsilon_{j}^{el}$), for the five frequencies of the orbital theory.}}
\end{center}
\end{figure}

\textcolor{black}{This lack of resonant amplification results in small absolute values of the obliquity amplitudes for the shell and the interior (always smaller than $10^{-3}$ degrees, see Fig. \ref{esei}). The absolute obliquity amplitudes of the interior (red regions in Fig.  \ref{esei}) are larger than the absolute obliquity amplitudes of the solid case (black line), which are in turn larger than those of the shell (blue regions): $|\varepsilon_{j,in}^{el}|>|\varepsilon_{j}^{el}|>|\varepsilon_{j,sh}^{el}|$. The absolute obliquity amplitudes increase with increasing ice shell thickness. For thin ice shells, $\varepsilon_{j,in}^{el}$ are close to $\varepsilon_{j}^{el}$, while for thick ice shells, $\varepsilon_{j,sh}^{el}$ tend to $\varepsilon_{j}^{el}$. This behavior suggests that the} layer that supports the main part of the aspherical mass distribution of the satellite tends to behave as if it was the entire satellite itself. This is a general behavior already observed for the  Galilean satellites in the absence of resonant amplification (Baland et al., 2012). Note that contrary to other authors (e.g. Nimmo and Spencer 2015), we do not consider that the presence of an internal global ocean is a sufficient condition for the obliquity of a satellite's shell being increased with respect to the \textcolor{black}{obliquity of a solid Enceladus}. A resonant amplification of a term of the solution is also needed, otherwise the shell obliquity is even smaller than the \textcolor{black}{obliquity of the solid case}.

\begin{figure}[!htb]
\begin{center}
\includegraphics[height=5cm]{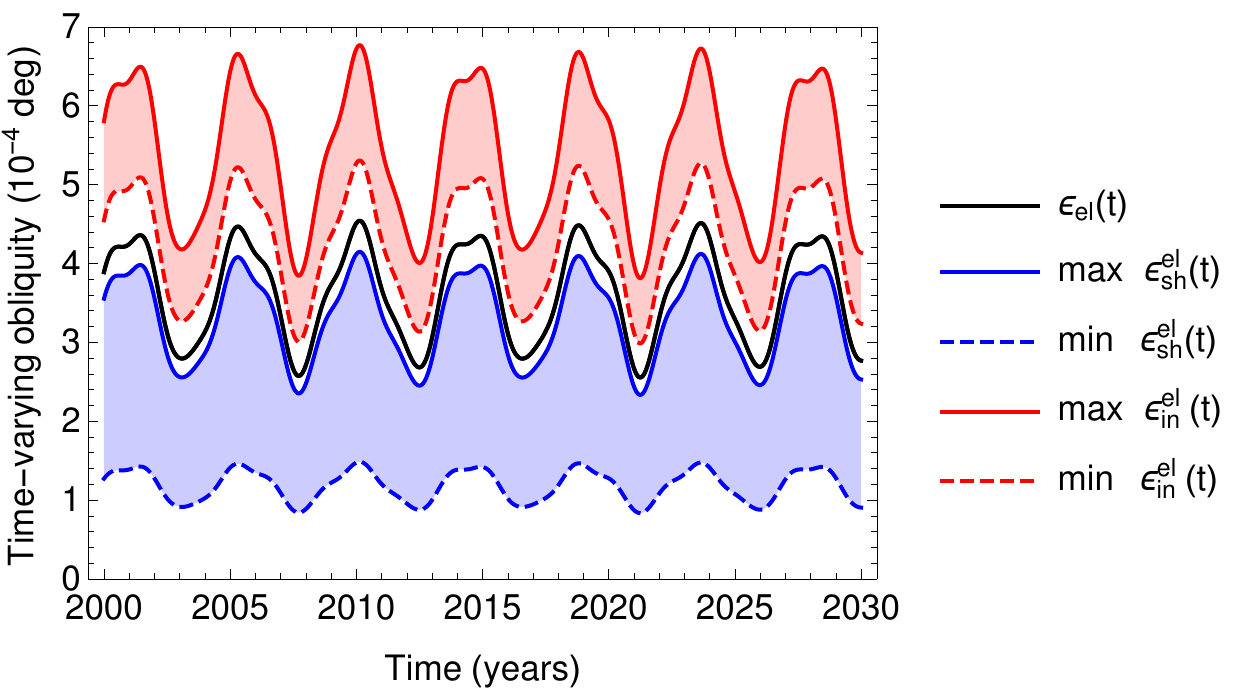}
 \caption{\label{Goblselastic} Ranges of the shell and interior obliquity $\varepsilon_{sh}^{el}(t)$ and $\varepsilon_{in}^{el}(t)$ of Enceladus over 30 years beginning on J2000, compared to the \textcolor{black}{obliquity of a solid and elastic Enceladus} $\varepsilon_{el}(t)$. The ranges are due to the diversity of the density profiles defined in Tab (2).}
\end{center}
\end{figure} 
\newpage

\begin{figure}[!htb]
\begin{center}
\includegraphics[width=12cm,trim = 40mm 35mm 0mm 40mm,clip]{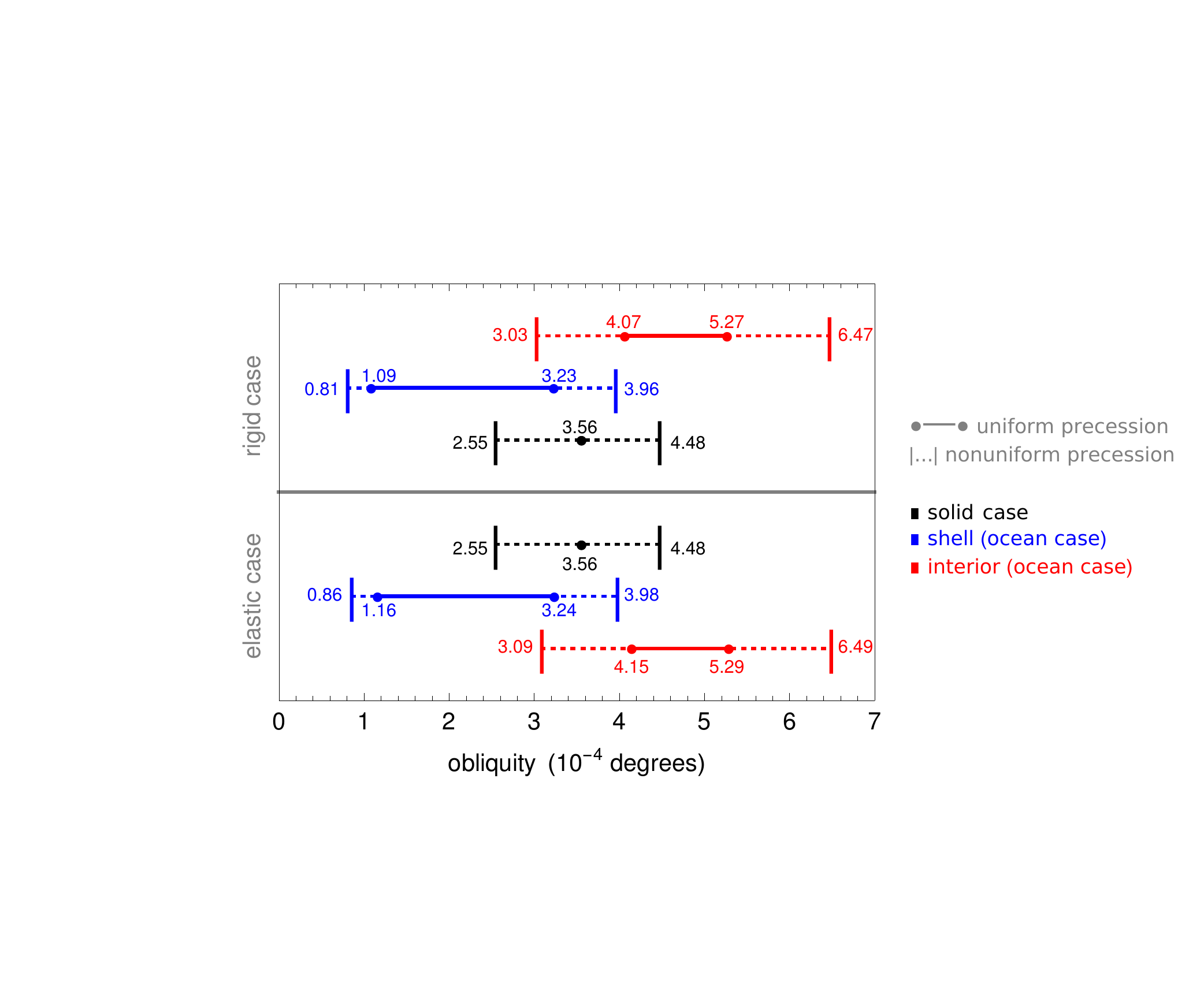}
 \caption{\label{upper}Ranges of the obliquity of Enceladus (in degrees), for the density profiles of Table \ref{tableEnceladus} and the eight different theoretical models for the Cassini state investigated in Sections \ref{section2} and \ref{section3}. Enceladus may be entirely solid or having an internal global ocean. The solid layers may be rigid or elastic. The orbital precession may be uniform or not.}
\end{center}
\end{figure}

If the orbital precession was uniform, the constant obliquity of the shell and of the interior \textcolor{black}{(computed from Eqs. (\ref{solution})-(\ref{solution2})} for $j=1$), $\varepsilon_{sh}^{el}$ and $\varepsilon_{in}^{el}$, would be in the ranges $[1.16,3.24]\times 10^{-4}$ degrees and $[4.15,5.29]\times 10^{-4}$ degrees, respectively, depending on the density profile. \textcolor{black}{However, since the precession is nonuniform, the actual obliquities of the shell $\varepsilon_{sh}^{el}(t)$ and of the interior $\varepsilon_{in}^{el}(t)$ (Eq. (\ref{actualobl})) vary with time in a similar way as the obliquity of an entirely solid and elastic Enceladus $\varepsilon_{el}(t)$. Depending on the density profile, the values for the obliquity of the interior are larger than for a solid Enceladus and in the range $[3.09,6.49]\times 10^{-4}$ degrees, while the values for the obliquity of the shell are smaller than for a solid Enceladus and in the range $[0.86,3.98]\times 10^{-4}$ degrees (see Figs. \ref{Goblselastic} and \ref{upper}).}

The effect of elasticity on the constant or on the time-varying obliquity is between \textcolor{black}{$0.5\%$} and $6.0\%$ for the shell and between $0.3\%$ and $1.9\%$ for the interior \textcolor{black}{(see Fig. \ref{upper})}. This is, as anticipated, a small effect. \textcolor{black}{As in the case of a solid moon, decreasing the ice rigidity increases the obliquity amplitudes. For $\mu_{sh}=1$ GPa, the obliquity amplitudes of the shell and of the interior are between $1.5\%$ and $15.6\%$ and between $1.1\%$ and $5.3\%$ larger than those of a rigid moon, respectively. In the limit case where the ice shell behaves as a fluid ($\mu_{sh}\rightarrow 0$), the obliquity amplitudes are about two times larger than the obliquity amplitudes for a moon with rigid solid layers.}

\section{Discussion}

\subsection{New theoretical estimate of the upper limit for Enceladus' obliquity}
\label{Section41}
The upper bound for Enceladus' obliquity was set to $0.0015^\circ$ by Chen and Nimmo (2011), using a model \textcolor{black}{for a solid and rigid moon} and a uniform precession. They increased this upper bound by about $20\%$ for a nonuniform precession. The ranges of Enceladus obliquity for the different theoretical models considered in this paper are given in Fig. \ref{upper}. \textcolor{black}{We showed in Section \ref{section2}, using the measured values of the gravity field coefficients, the quasi-periodic decomposition for the orbital precession, and reasonable estimates of the ice and interior rigidities ($3.3$ GPa and $100$ GPa, respectively), that the upper limit for a solid Enceladus is even smaller than Chen and Nimmo's estimate: $4.48\times 10^{-4}$ degrees, whether the satellite is rigid or elastic}. When the presence of an internal ocean is taken into account, the upper bound for the obliquity of a rigid/elastic shell decreases \textcolor{black}{by $13.1/12.6\%$ to reach $3.96/3.98\times 10^{-4}$ degrees}. This corresponds to about two meters at the surface of the satellite. \textcolor{black}{For an ice shell deforming as a fluid layer (unrealistic limit case), the upper limit on the obliquity of Enceladus would be about $9\times 10^{-4}$ degrees ($4$ meters), which is still quite small.}

A very accurate measurement of the obliquity would be needed to detect the ocean, especially if the shell is thick. For instance, an accuracy of the order of $5\times 10^{-5}$ degrees would be needed to detect an ocean below a $25$ km thick shell, corresponding to an error of the order of $0.20$ m on the location of the spin axis at the surface. This accuracy level is well below the resolution of the Cassini images used by Giese (2014) and ranging from $112$m/pixel to $1217$m/pixel. \textcolor{black}{Therefore, control points calculations are insufficient to detect the ocean or measure the obliquity of Enceladus.}

The ocean model fails to predict larger values for the surface obliquity than the solid model. The obliquity is about two orders of magnitude smaller than the obliquity needed to explain the observed heat flux of Enceladus, according to Tyler (2009, 2011). Therefore, as Chen and Nimmo (2011), we conclude that obliquity tides are unlikely to be the source of the large heat flow and geophysical activity of Enceladus.

\subsection{Deviation from hydrostatic equilibrium}

From the observed gravity field, Iess et al. (2014) showed that $C_{20}/C_{22}=-3.51\pm0.05$. This is not consistent with the canonical hydrostatic relation $C_{20}/C_{22}=-\frac{10}{3}$, which is a necessary condition for a synchronous satellite to be in hydrostatic equilibrium. The existence of a significant degree-three signal in the gravity field is another evidence of a departure from hydrostatic equilibrium.  McKinnon (2015) showed that for a rapid rotator as Enceladus (period of rotation of 1.37 days) the hydrostatic relation is closer to 3.25, which is even further from the observations. Enceladus is definitely not in hydrostatic equilibrium. Therefore, the mean moment of inertia of Enceladus cannot be computed straightforwardly from the observed gravity field and Radau's equation. One needs to separate the hydrostatic and non-hydrostatic components of the measured gravity field first. Assuming that the interior (the \textcolor{black}{ice/silicates} core) of Enceladus is in hydrostatic equilibrium and that the floating shell is isostatic and of variable thickness (Airy isostasy), Iess et al. (2014) found that $I/M_e R^2=0.335\pm0.005$ at the $2\sigma$ level, which is the value we used in Section \ref{section2} to approximate the ratio $C/M_e R^2$ and in section 3 to constrain the range of density profiles of Enceladus in Table \ref{tableEnceladus}.

Since Enceladus is not in hydrostatic equilibrium, it is not surprising that the reference ellipsoid computed from the observed gravity field is not consistent with the measured topography of Nimmo et al. (2011). Expected hydrostatic values for the radii $a>b>c$ of the surface ellipsoid are $a=255.7$ km, $b=251.0$ km, and $c=249.3$ km, while the ellipsoid fitted to the topography measurements are characterized by $a=256.7$ km, $b=251.2$ km, and $c=248.2$ km. \textcolor{black}{Therefore}, the polar and equatorial static flattenings of the surface, of the ice shell-ocean interface, and of the ocean-interior interface deviate from the hydrostatic flattenings computed in section (\ref{profiles}) and used to derive the solution for the spin precession in section (\ref{computation}). The flattenings of the ellipsoid fitted to the observed surface topography are \textcolor{black}{derived from Nimmo et al. (2011): $\alpha_{sh}=0.0229$ and $\beta_{sh}=0.0218$.}

\textcolor{black}{We compute the non-hydrostatic flattenings of the shell-ocean and ocean-interior interface's that are consistent with the observed degree-two gravity coefficients $C_{20}$ and $C_{22}$ and the ellipsoid fitted to the measured surface topography, under the assumptions that the interior is in hydrostatic equilibrium while the shell is in a quasi-isostatic equilibrium state (meaning an Airy-type compensation through ice-shell thickness variations, allowed by a density contrast between the ocean and the shell). For the interior in hydrostatic equilibrium, we can write the following relations between its flattenings $\alpha_{in}$ and $\beta_{in}$ and the flattenings of the other layers (see Baland et al., 2014):} 
\begin{eqnarray}
-\frac{2}{3}\alpha_{in} R_{in}&=&\frac{\Phi_{in}^{20}(R_{in})}{g(R_{in})},\\
\frac{1}{6}\beta_{in} R_{in}&=&\frac{\Phi_{in}^{22}(R_{in})}{g(R_{in})},
\end{eqnarray}
with 
\begin{eqnarray}
\nonumber \Phi_{in}^{20}(R_{in})&=&-\frac{5}{6} n^2 R_{in}^2-\frac{8\pi G}{ 15}\left[\alpha_{sh}\rho_{sh} R_{in}^2+\alpha_o(\rho_o-\rho_{sh})R_{in}^2\right.\\
\label{69}&&\left.-\alpha_{in}\rho_o R_{in}^2+\alpha_{in}\rho_{in}R_{in}^2\right],\\
\nonumber \Phi_{in}^{22}(r_{in})&=&\frac{1}{4} n^2 r_{in}^2+\frac{2\pi G}{15}\left[\beta_{sh}\rho_{sh} R_{in}^2+\beta_o(\rho_o-\rho_{sh})R_{in}^2\right.\\
\label{70}&&\left.-\beta_{in}\rho_o R_{in}^2+\beta_{in}\rho_{in}R_{in}^2\right],
\end{eqnarray}
with $g(R_{in})=G M(R_{in})/R_{in}^2$, the gravitational acceleration at level $R_{in}$. $M(R_{in})$ is the mass of the material located under the mean level $R_{in}$.
The observed second-degree gravity coefficients can be expressed as a function of the layers' flattenings \textcolor{black}{(see Baland et al., 2014)}:
\begin{eqnarray}
\nonumber C_{20}&=&-\frac{1}{M_e R^2}\frac{8\pi}{15}\left[\rho_{sh}(R^5\alpha_{sh}-R_o^5\alpha_o)+\rho_o(R_o^5\alpha_o-R_{in}^5\alpha_{in})\right.\\
\label{71}&&\left.+\rho_{in}R_{in}^5\alpha_{in}\right],\\
\nonumber C_{22}&=&\frac{1}{4 M_e R^2}\frac{8\pi}{15}\left[\rho_{sh}(R^5\beta_{sh}-R_o^5\beta_o)+\rho_o(R_o^5\beta_o-R_{in}^5\beta_{in})\right.\\
\label{72}&&\left.+\rho_{in}R_{in}^5\beta_{in}\right].
\end{eqnarray}

\begin{figure}[!htb]
\begin{center}
\includegraphics[width=14cm]{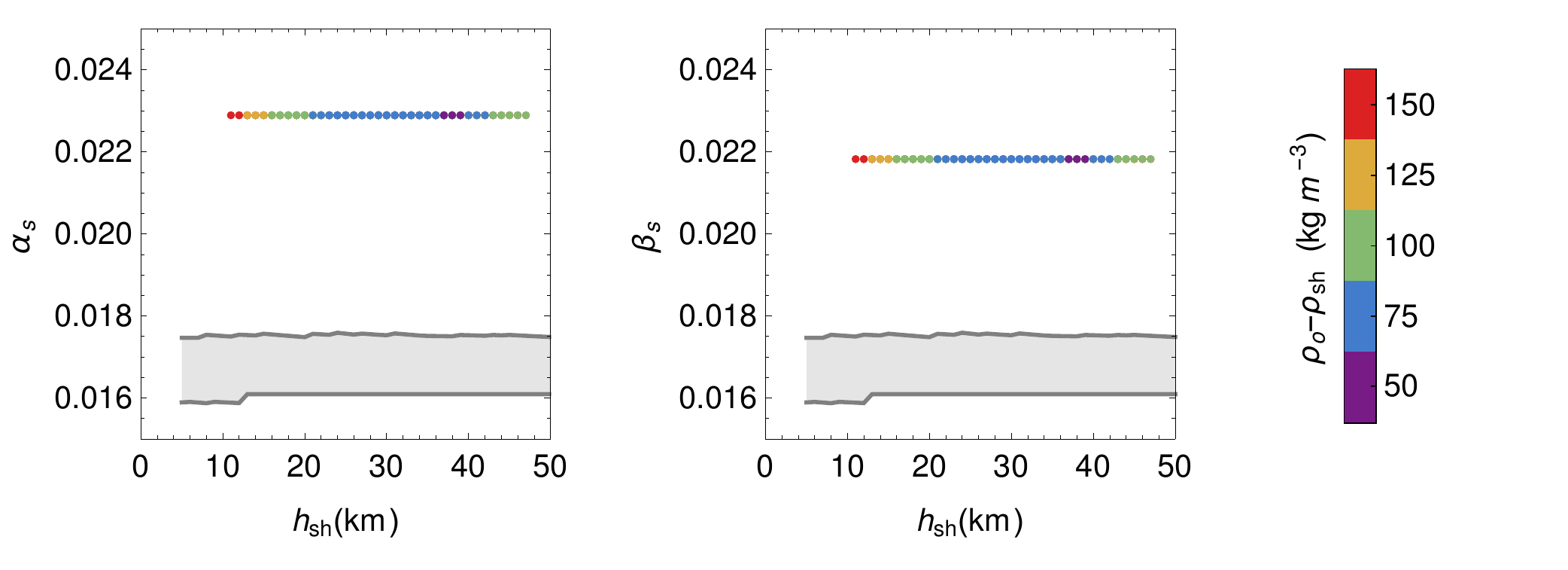}
\includegraphics[width=14cm]{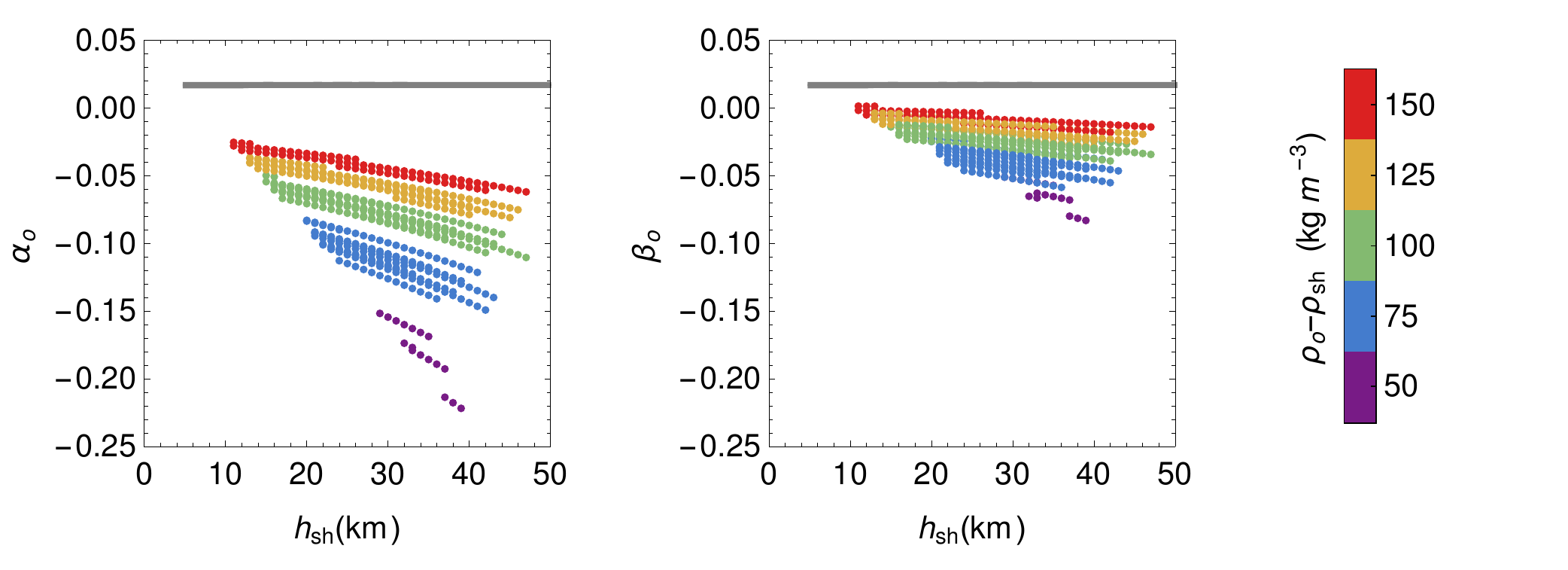}
\includegraphics[width=14cm]{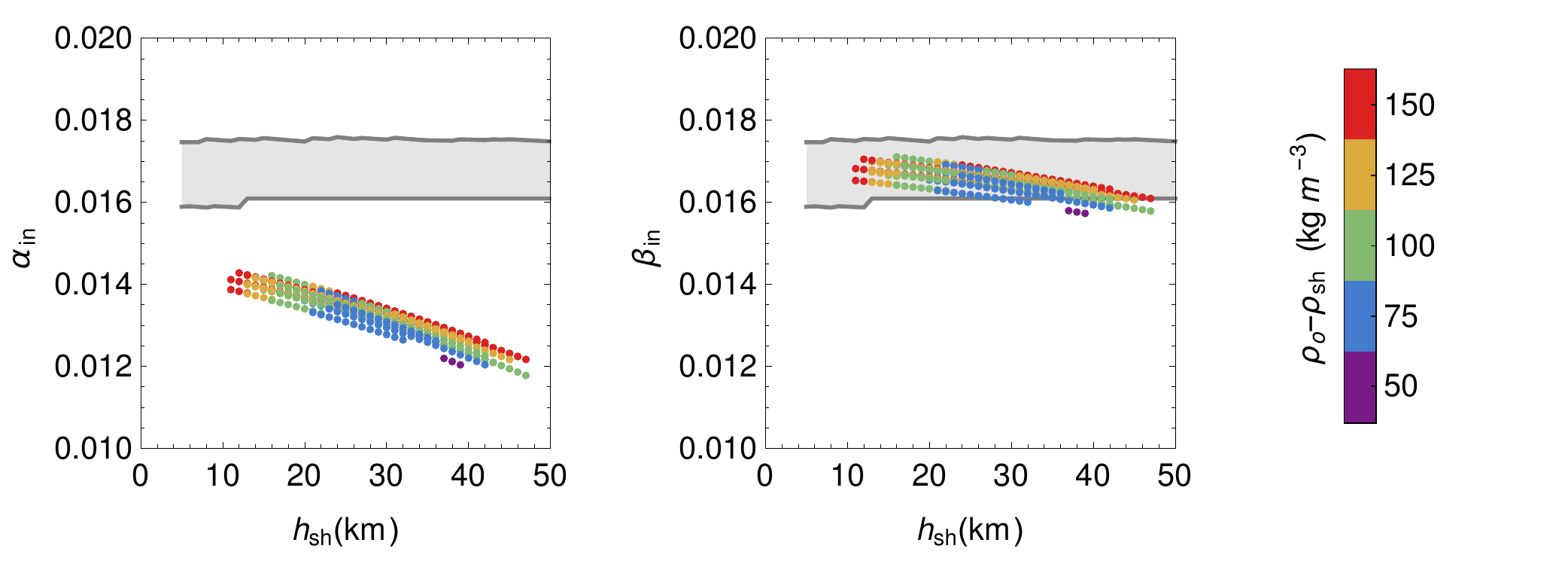}
 \caption{\label{alphabeta}Polar ($\alpha$) and equatorial ($\beta$) flattenings of the shell, ocean, and interior of Enceladus, as a function of the ice shell thickness $h_{sh}$, for the density profiles defined in Table \ref{tableEnceladus}. In gray, the ranges of hydrostatic flattenings used in section (\ref{computation}). In markers colored according to the density difference $(\rho_o-\rho_{sh})$, the non-hydrostatic flattenings corresponding to the measured topography and gravity fields.  }
\end{center}
\end{figure}

Eqs. (\ref{69})-(\ref{72}) form a system of four equations with four unknowns: $\alpha_o$, $\alpha_{in}$, $\beta_o$, and $\beta_{in}$. \textcolor{black}{For $\rho_o=\rho_{sh}$, an isostatic compensation at the shell-ocean interface is impossible. Therefore, the solution can be computed only for the 1869 profiles, out of the 2191 density profiles presented in Table \ref{tableEnceladus}, which have a density contrast $\Delta \rho=(\rho_o-\rho_{sh})$ of $25, 50, \dots, 150$ km. The solution for the flattenings allows to estimate the thickness of the shell and of the ocean for any longitude and latitude. The quasi-isostatic compensation leads to a thinner shell and a thicker ocean at the poles, and to a thicker shell and a thinner ocean in the direction to Saturn than in the hydrostatic case. However, some solutions are physically unrealistic. We find negative shell thicknesses at the poles (up to $-40$ km) and negative ocean thicknesses in the direction to Saturn (up to $-65$ km), for density profiles with small ($h_{sh}<8$ km) and large ($h_{sh}>48$ km) mean shell thickness, respectively, and/or for density profiles with a low density contrast $\Delta \rho$ of $25$ km. These unrealistic solutions are the direct consequence of the hypothesis of quasi-isostatic compensation at the ocean-shell interface, that implies that the shell-ocean interface is significantly less flattened than the shell's surface or even flattened in the perpendicular direction. In the latter case, for a small mean shell thickness, the ocean-shell interface will easily cross the shell's surface in the polar direction, while for a large mean shell thickness, the mean ocean thickness will be small and the shell-ocean interface will easily cross the core's surface in the direction to Saturn. For intermediate mean shell thickness, a low density contrast between the shell and the ocean produces similar effects, because the isostasy will lead to negative $\alpha_o$ and $\beta_o$ that are an order of magnitude larger than the hydrostatic flattenings. Therefore, in Fig. \ref{alphabeta}, we present the solution for the flattenings, as a function of the mean shell thickness $h_{sh}$ and of the density contrast $\Delta \rho$, for the subset of $555$ density profiles which have a shell and ocean thickness of at least 2.5 and 1 km at all directions, half the minimal value considered here for the mean thickness of the shell and of the ocean, respectively. The surface flattenings have the same value for all density profiles (see top panels of in Fig. \ref{alphabeta}). Ocean (middle panels) and interior (bottom panels) flattenings may differ depending on the density profile, and decrease with decreasing $h_{sh}$. $\alpha_{in}$ differs more with respect to the hydrostatic flattenings than $\beta_{in}$. This is because the difference between the actual surface flattenings and the hydrostatic one is larger for the polar flattenings than for the equatorial ones (see top panels in Fig. \ref{alphabeta}).}

Using the non-hydrostatic flattenings instead of the hydrostatic flattenings to derive the solution for the spin precession, we find that \textcolor{black}{the non-hydrostatic behavior of the shell has a small effect on the obliquity of the shell, which is now in the range $[1.06,3.35]\times 10^{-4}$ deg. The possible difference in ocean thickness between the two poles, due to the south pole depression, is unlikely to change significantly the outcome.} 

\subsection{Implications of elasticity for other satellites}

\textcolor{black}{The effect of elastic deformations on the obliquity of the surface layer is very limited in the case of Enceladus (from $0.3\%$ to $6.0\%$, see Section \ref{computation}).} But what about other satellites? As a preliminary assessment of the effect of elasticity on the Galilean satellites and on Titan, we compare \textcolor{black}{in Table \ref{othersat}} their rigid and elastic solid free periods, considering reasonable estimates for their solid tidal Love number $k_2$: $0.04$ to $0.8$ for Io, depending on the viscosity of the astenosphere (Lainey et al., 2009), $0.015$ for Europa (Moore and Schubert, 2000), $0.067$ for Ganymede, Callisto, and Titan (Moore and Schubert, 2003).

\begin{table}[htdp]
\begin{center}
\begin{tabular}{lccccc}
\hline
& $k_2$ & $T_f$ (years) & $T_f^{el}$ (years)& $\Delta T_f$&\\
\hline
Enc. &  $0.00015$ & $0.098$ & $0.098$&$0.01\%$\\
Io   &  $0.04-0.8$ & $0.410$& $[0.420,0.759]$&$[2.3\%,46\%]$\\
Eur. &  $0.015$ & $3.211$& $3.246$&$1\%$\\
Gan. &  $0.067$& $19.967$& $21.298$&$6\%$ \\
Cal. &  $0.067$ & $203.076$ & $213.029$&$4.7\%$\\
Titan&  $0.067$ & $191.943$ & $202.295$&$5.1\%$\\
&  $0.589\pm0.075$ (*) &  & $[316.013,389.487]$&$[35.6\%,56.6\%]$\\
\hline
\end{tabular}\\
\end{center}
\caption{\label{othersat}Effect of elastic deformations on the free period of Enceladus, the Galilean satellites, and Titan, considered as solid bodies uniformly precessing. $k_2$ are the \textcolor{black}{estimates of the Love numbers for solid moons}, except for (*), which is the measured value for Titan. $T_f$ and $T_f^{el}$ are the \textcolor{black}{free periods for rigid and elastic moons, respectively, while $\Delta T_f=(T_f^{el}-T_f)/T_f^{el}$ is relative difference of these free periods.} }
\end{table}

\textcolor{black}{Since the $k_2$ of other satellites is at least two orders of magnitude larger than for Enceladus (see Eq. (\ref{29}) and Table \ref{othersat}), the effect of elastic deformations on the free period is expected to be larger for the other satellites as well, as shown in Table \ref{othersat}. The increase of the free period} for the Galilean satellites ranges from $1\%$ (Europa) to $46\%$ (Io with a low viscosity astenosphere). Baland et al. (2012) showed that for Io in the rigid case, there was a slight resonant amplification of the fifth term of the solution which corresponds to a forcing period of $-0.68$ years. With a low viscosity astenosphere, the free period can get very close to the forcing period \textcolor{black}{($T_f^{el}$ is between 0.42 and 0.76 years, see Table \ref{othersat}). This would lead} to a larger obliquity than predicted in Baland et al. (2012). For Europa, Ganymede, and Callisto, using theoretical $k_2$ values corresponding to a solid case, \textcolor{black}{as done in Table \ref{othersat}}, may lead to underestimate the effect of elasticity on the spin precession, since these satellites likely harbor an internal global ocean. \textcolor{black}{As shown here for Enceladus, the effect of elasticity is about two orders of magnitude larger if an ocean is present in the interior, as it leads to larger tidal deformations. The same applies to Titan.} 

The \textcolor{black}{constant obliquity of a solid and rigid Titan} for a uniform precession is of $0.12^\circ$, which is inconsistent with the measured value of $0.32^\circ\pm0.02^\circ$ (Bills and Nimmo, 2008; Baland et al., 2011). If we assume that Titan precesses as a solid body but deforms elastically according to the measured $k_2=0.589\pm0.075$, which is an order of magnitude larger than the \textcolor{black}{theoretical estimate for a solid Titan, we find a free period} about $45\%$ larger than the rigid one. This brings the free period closer to the forcing period ($-703.51$ years) than with the solid $k_2$, leading to a more important resonant amplification of the obliquity which is now in the range $[0.26,0.40]$ degrees and is consistent with the observed rotation state of Stiles et al. (2008, 2010). Baland et al. (2011, 2014) explained the measured obliquity by a resonant amplification of the solution for a Titan with an internal ocean and rigid solid layers. However, the appropriate resonant amplification was met for only a very few density profiles among the whole range of density profiles considered. The preliminary results presented here indicate that more density profiles would meet the appropriate resonant amplification, which would influence the interpretation of the measured obliquity of Titan in terms of internal structure of Baland et al. (2014). \textcolor{black}{The chosen values for the rigidity of the solid layers (especially for the ice shell) would also influence the constraints placed on the interior.}

\section{Conclusions}

We have developed a new model for the multi-frequency Cassini state of a synchronous satellite which harbors an internal global ocean and deforms elastically. The model takes into account the external torque exerted by Saturn on each layer of the satellite and the internal gravitational and pressure torques induced by the presence of the liquid layer. We applied the model to Enceladus, which is known for its strong heat flux, possibly caused by obliquity tides, according to Tyler (2009, 2011). We found that, for Enceladus with a subsurface ocean, the surface has a maximal obliquity of $4.0\times 10^{-4}$ degrees, taking into account the nonuniform precession of the orbit.

This upper limit is smaller than the previous estimate by Chen and Nimmo (2011) and, unfortunately, than Tyler's requirement ($0.05^\circ-0.1^\circ$). Therefore, the strong heat flux of Enceladus is more likely the result of another process, for example the eccentricity tides. The upper limit is also well below the achievable accuracy of Cassini images. Therefore, control point calculations will not have enough sensitivity to measure the obliquity of Enceladus, let alone to constrain its interior from an obliquity measurement. 

Although this study concludes with a negative result concerning Enceladus, it opens interesting perspectives for other satellites like the Galilean satellites and Titan. Larger satellites with larger tidal Love numbers will be more affected than Enceladus by the effect of elastic deformations. For instance, the rotation state of Io can be strongly affected if its astenosphere is of low-viscosity, while the ocean of Europa, Ganymede, Callisto, and Titan would lead to large elastic deformations with a significant impact on the rotation state. This could alter the prediction of Baland et al. (2012) for the obliquity of the Galilean satellites, and the interpretation of the measured obliquity of Titan in terms of internal structure of Baland et al. (2014).  Studying the combined effect of elasticity and of an internal ocean on the spin precession of the Galilean satellites and of Titan was beyond the scope of the present study, but will be an important question to address in the future. The case of Titan is a very important one, since both its obliquity and tidal Love number have been measured and are larger than usually expected. 

\section*{Acknowledgments}

\textcolor{black}{The authors thank Mikael Beuthe for fruitful discussions on the Love number computation and two anonymous reviewers whose valuable comments improve the paper. R.-M. Baland is funded by a FSR (Fonds sp\'{e}cial de recherche) grant from UCL. The research leading to these results has received funding from the Belgian PRODEX program managed by the European Space Agency in collaboration with the Belgian Federal Science Policy Office. }

\appendix

\section{Angular momentum equation for a solid elastic Enceladus}

We here provide a demonstration of Eq. (\ref{elasticcase}) for the angular momentum equation governing the spin precession of an elastic solid satellite. We extend the demonstration given in appendix A of Baland et al. (2012) for a rigid solid satellite.

\subsection{Time-variable gravity field}

\begin{figure}[!htb]
\begin{center}
\includegraphics[width=12cm]{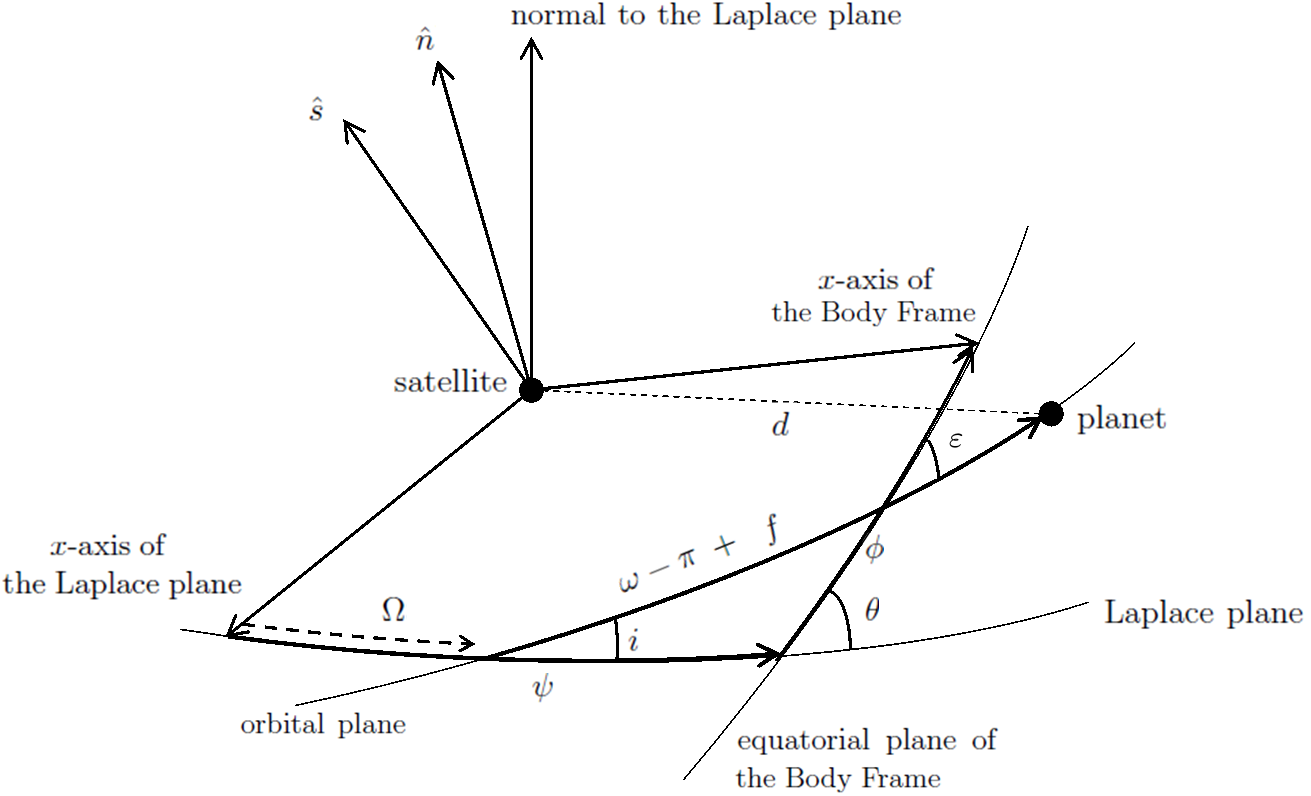}
\caption{\label{FigA1} Orientation of the Body Frame of the satellite with respect to its orbital plane and to its Laplace plane, from Baland et al. (2012). }
\end{center}
\end{figure}

The external degree-two gravitational potential of Enceladus $ V^{l=2}$, exerted at a radial distance $r$, colatitude $\varphi$, and longitude $\lambda$, is the result of deformations induced by the tidal and centrifugal degree-two potentials $V_t$ and $V_c$:
\begin{eqnarray}
 \nonumber&& V^{l=2}(r,\varphi,\lambda)\\
 \nonumber&&=-\frac{GM_e}{r}\left(\frac{R}{r}\right)^2\sum_{m=0}^{2}(C_{2m} \cos m\lambda + S_{2m} \sin m\lambda)P^m_2(\cos \varphi)\\
 \nonumber&&=k_f \left(\frac{R}{r}\right)^3 (V_t^{stat}(R,\varphi,\lambda)+V_c^{stat}(R,\varphi,\lambda))\\
 \label{A2}&&\quad +k_2 \left(\frac{R}{r}\right)^3  (V_t^{peri}(R,\varphi,\lambda)+V_c^{peri}(R,\varphi,\lambda)),
\end{eqnarray}
where $M_e$ and $R$ are the mass and the mean radius of Enceladus, $G$ is the universal gravitational constant, $C_{2m}$ and $S_{2m}$ are second-degree gravity field coefficients, $k_f$ is the fluid Love number and $k_2$ is the tidal Love number, $P^m_2$ is the Legendre function of degree two and order $m$. The tidal and centrifugal potentials have been divided into their static and periodic parts, which are written, at first order in orbital eccentricity $e$, orbital inclination $i$, satellite equatorial plane inclination with respect to the Laplace plane $\theta$, and libration in longitude angle $\gamma$ as:
\begin{eqnarray}
 V_t^{stat}(r,\varphi,\lambda)&=&-q_t \frac{GM_e}{6R}\left(\frac{r}{R}\right)^2\left[P^0_2(\cos \varphi)-\frac{1}{2}P_2^2(\cos \varphi)\cos 2\lambda\right],\\
 V_c^{stat}(r,\varphi,\lambda)&=&q_r \frac{GM_e}{3 R}\left(\frac{r}{R}\right)^2 P^0_2(\cos \varphi),\\
 \nonumber V_t^{peri}(r,\varphi,\lambda)&=&q_t \frac{GM_e}{3R}\left(\frac{r}{R}\right)^2 (3 e\cos M)\\
 \nonumber &&\left[-\frac{1}{2}P^0_2(\cos \varphi)+\frac{1}{4}P_2^2(\cos \varphi)\cos 2\lambda\right]\\
 \nonumber &&+q_t \frac{GM_e}{3R}\left(\frac{r}{R}\right)^2 \left[\frac{1}{4}P_2^2(\cos \varphi)\sin 2\lambda \right.\\
 \nonumber &&\left.(4 e \sin M-2 \gamma)-P_2^1(\cos \varphi)\cos \lambda\right.\\
\label{A4} &&\left.\lbrace\theta \sin(\omega+M+\Omega-\psi)-i\sin(\omega+M)\rbrace \right],\\
 V_c^{peri}(r,\varphi,\lambda)&=&2 \frac{\dot\gamma}{n}q_r \frac{GM_e}{3 R}\left(\frac{r}{R}\right)^2 P^0_2(\cos \varphi),
\end{eqnarray}
with
\begin{equation} 
 q_t=-3\frac{GM_p}{GM_e}\left(\frac{R}{a}\right)^3\quad \textrm{and}\quad q_r=\frac{n^2R^3}{GM_e}.
\end{equation}
$M_{p}$ is the mass of the planet Saturn, $a$ is the semi-major axis of Enceladus, and $n$ is the mean motion of Enceladus. $q_t=-3q_r$ because of Kepler's third law ($GM_p=n^2a^3$). $\Omega$, $\omega$, and $M$ are \textcolor{black}{the longitude of the ascending node, the argument of pericenter, and the mean anomaly of the satellite's orbit with respect to the Laplace plane (the argument of the pericenter of the planet seen as in orbit around the satellite is then $\omega-\pi$). $\psi$ is the ascending node longitude} of the equatorial plane of the satellite with respect to the Laplace plane (see Fig. \ref{FigA1}). Note that we have neglected the polar motion/wobble here. Therefore the spin vector of Enceladus is aligned with the polar axis. 

\begin{figure}[!htb]
\begin{center}
\includegraphics[width=13cm]{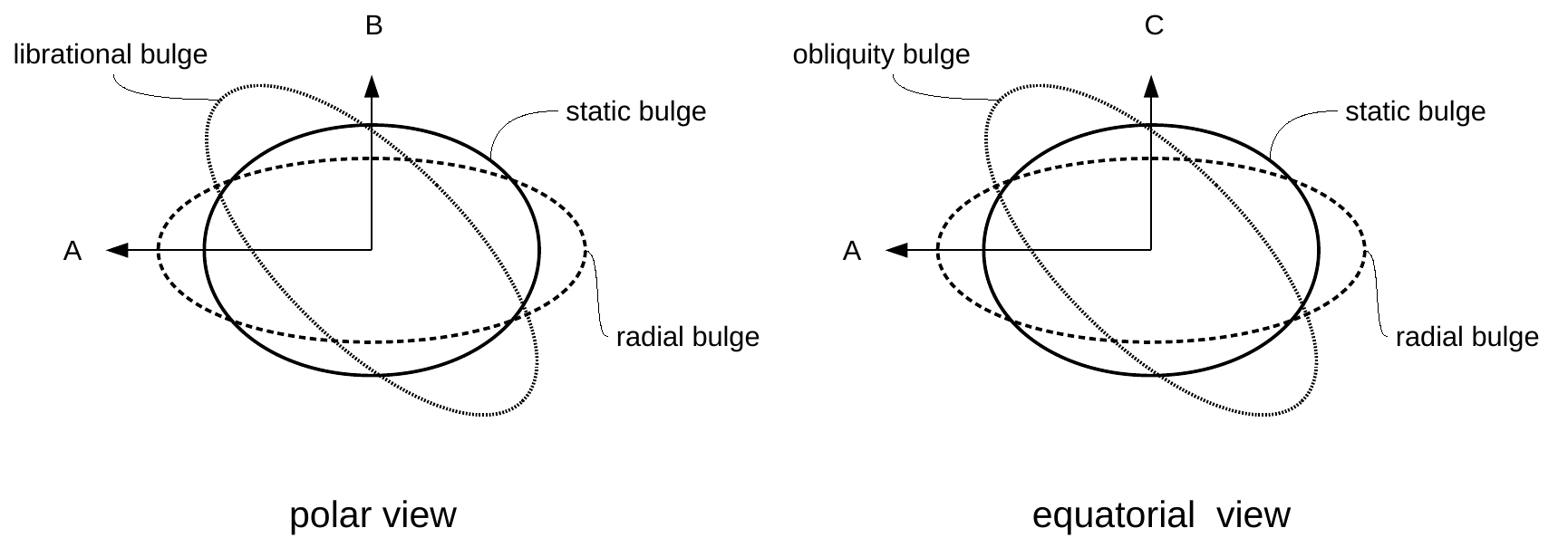}
\caption{\label{FigA2}Orientation of the periodic bulges with respect to the static bulge, seen from the equatorial plane (defined by the A and B axes) and from the plane defined by the shortest and largest axes of the satellite (the A and C axes). The radial, librational and obliquity bulges vary in amplitude over time. }
\end{center}
\end{figure}

Using Eq. (\ref{A2}), the gravity coefficients are given, at first order in $e, i, \theta$ and $\gamma$, by
\begin{eqnarray}
  \label{C20}C_{20}&=&k_f\left(\frac{-2q_r+q_t}{6}\right)-\frac{3}{2} k_2\,e\,q_r \cos M-\frac{2}{3} k_2  q_r \frac{\dot \gamma}{n},\\
    \label{C22}C_{22}&=&-k_f \frac{q_t}{12}+\frac{3}{4} k_2\,e\,q_r \cos M,\\
  S_{22}&=& k_2 q_r (e \sin M - \gamma/2),\\
 \label{C21} C_{21}&=& k_2 q_r [i \sin{(\omega+M)}-\theta \sin{(\omega+M+\Omega-\psi)}],\\
  \label{S21}S_{21}&=&0.
\end{eqnarray}
The non static terms in Eqs. (\ref{C20})-(\ref{S21}) represent the elastic tidal bulge which can be divided into in three components (radial, librational, and obliquity) having fixed directions with respect to the static bulge, but time-varying amplitudes (see Fig.\ref{FigA2}). The terms proportional to $e$ in $C_{20}$ and $C_{22}$ are due to the radial eccentricity tides, while the term proportional to $\dot \gamma$ in $C_{20}$ is due to the librations changing the rotation rate in the centrifugal potential. These terms form the ``radial bulge'', which is aligned with the static bulge. The term proportional to $e$ in $S_{22}$ corresponds to the ``eccentricity librational tidal bulge'', which differs by $45^\circ$ from the orientation of the static bulge, in the equatorial plane.  This eccentricity librational bulge is counteracted by a term proportional to $\gamma$, and they can be together called ``librational tides''. The term proportional to $i$ and $\theta$ \textcolor{black}{in $C_{21}$ corresponds} to the ``obliquity tidal bulge'', which differs by $45^\circ$ from the orientation of the static bulge, in the plane defined by the moments of inertia $A$ and $C$. \textcolor{black}{In the Cassini state, it is often assumed that $\psi=\Omega$ and that $\theta-i= \varepsilon$, where $\varepsilon$ is the obliquity (see e.g. Tyler, 2011 or Appendix C of Beuthe, 2013). Here we chose not to make these approximations directly, because they make sense only posteriori, once the solution of the angular momentum averaged over the orbital period has been obtained, and for a satellite with a uniform orbital precession (see Henrard, 2005 and Section \ref{section23}).}

\subsection{External Torque}

With the inertia tensor of the solid Enceladus defined as
\begin{eqnarray}\label{I}
\bar{\bar I}=\left(
\begin{array}{ccc}
 A  &- F & -E \\
 - F& B & - D \\
- E & - D& C
\end{array}
\right),
\end{eqnarray}
the expression of the torque in the frame attached to the mean principal axes of inertia of the satellite (Body Frame) is (adapted from Murray and Dermott 1999, Eqs 5.43-5.45):
\begin{equation}
\vec \Gamma_{BF}=\frac{3 n^2 a^3}{d^5}\left(
\begin{array}{c}
 (C- B) Y Z - D(Y^2-Z^2)-E\, XY+ F \,XZ \\
 (A- C) Z X +D X Y+ E (X^2-Z^2)- F\, YZ \\
  (B- A) X Y -D X Z+ E \,YZ- F(X^2-Y^2)
\end{array}\right),
\end{equation}
with $d$ the distance between the satellite and the planet, and $(X,Y,Z)$ the position of the planet in the Body Frame in Cartesian coordinates. 

The moments of inertia are related to the time-varying coefficients $C_{lm}$ and $S_{lm}$ of the gravitational potential of the satellite in the following way:
\begin{eqnarray}
 \label{A14} C-A&=&M_e R^2 (-C_{20}+2C_{22}),\\
 C-B&=&M_e R^2 (-C_{20}-2C_{22}),\\
 \label{A16} B-A&=&4 M_e R^2 C_{22}, \\ 
 D&=&-M_e R^2  S_{21},\\
 \label{A18} E&=&-M_e R^2  C_{21},\\
\label{A19} F&=& 2 M_e R^2  S_{22},
\end{eqnarray}
where $C_{lm}$ and $S_{lm}$ can be replaced by Eqs. (\ref{C20})-(\ref{S21}). The diagonal moments of inertia $A<B<C$ have non zero mean values ($\bar A, \bar B, \bar C$), while the off-diagonal elements $E$ and $F$ vary periodically in time about zero. $D=0$ since $S_{21}=0$.  

The coordinates of the planet can be expressed as
\begin{equation} 
\left(
\begin{array}{c}
 X \\
 Y \\
 Z
\end{array}\right)=R_z(\phi).R_x(\theta).R_z(\psi-\Omega).R_x(-i).R_z(-\omega+\pi-f)\left(
\begin{array}{c}
d \\
 0 \\
 0
\end{array}\right),
\end{equation}
with $f$ the true anomaly of the satellite and the rotation matrices defined as
\begin{eqnarray}\label{RxRz}
R_x(\theta) = \left(
\begin{array}{ccc}
 1 & 0 & 0 \\
 0 & \cos{\theta} & \sin{\theta} \\
 0 & -\sin{\theta} & \cos{\theta}
\end{array}
\right) \, \textrm{and}\,  R_z(\theta) = \left(
\begin{array}{ccc}
 \cos{\theta} & \sin{\theta} & 0 \\
 -\sin{\theta} & \cos{\theta} & 0 \\
 0 & 0 & 1
\end{array}
\right).
\end{eqnarray}

Correct up to the first order in orbital eccentricity $e$,
\begin{eqnarray}
f&\simeq&M+2e \sin{M},\\
d&\simeq&a-a\, e \cos{M}.
\end{eqnarray}
Because of the synchronicity of the rotation with the orbital revolution, the Euler angle $\phi\simeq -\psi+\Omega+\omega-\pi+M+\gamma$ (e.g. Peale 1969), correct up to the first order in the small angles $i$, $\theta$, and $\gamma$. Therefore, we have that 
\begin{eqnarray}\label{A7}
\nonumber &&\vec \Gamma_{BF}\simeq 3 n^2 \\
&&\left(\begin{array}{c}
 0 \\
 \left[(\bar C-\bar A)-k_2 q_r M_e R^2\right] [i \sin{(\omega+M)}-\theta \sin{(\omega+M+\Omega-\psi)}] \\
 \left[(\bar B-\bar A)-k_2 q_r M_e R^2\right] (2 e \sin M -\gamma)
\end{array}\right).
\end{eqnarray}
At first order, the effect of elastic deformations on the torque is solely due to the off-diagonal elements of the inertia tensor $E$ and $F$. The effect of the time variations of the diagonal elements is only a second-order effect. 
See also Van Hoolst et al. (2013) for a shorter demonstration for the z-component only, and Coyette et al. (2016) for a similar demonstration taking into account a possible polar motion neglected here. The effect of elastic deformations results in a multiplication of the parts of $(\bar C-\bar A)$ and $(\bar B-\bar A)$ related to the static tidal potential by $\left(\frac{k_f-k_2}{k_f}\right)$:
\begin{eqnarray}
(\bar B-\bar A)- k_2  q_r M_e R^2 &=& (\bar B-\bar A)_t\left(\frac{k_f-k_2}{k_f}\right),\\
(\bar C-\bar A)- k_2 M_e  q_r R^2 &=& (\bar C-\bar A)_c +(\bar C-\bar A)_t\left(\frac{k_f-k_2}{k_f}\right).
\end{eqnarray}
In other words, the effect of the periodic part of the tidal potential is to counteract the effect of the static part, through the librational and obliquity tides arising in the coefficients $S_{22}$ and $C_{21}$, respectively. The torque for the rigid case, to be used in Eq. (\ref{rigidcase}), \textcolor{black}{can obtained by  setting $k_2=0$ in Eq. (\ref{A7})}. 

The bars over the static part of the principal moment of inertia are now omitted for simplicity. The periodic part of a moment of inertia will be preceded by $\delta$ to make the distinction with respect to the static values. For instance, $C$ and $E$ will be now denoted $C+\delta C$ and $\delta E$, respectively, where $C$ is the static polar moment of inertia and $\delta C$ and $\delta E$ are \textcolor{black}{inertia increments} due to the tidal and centrifugal elastic deformations. \textcolor{black}{The expression for $\delta C$ is obtained from Eqs. (\ref{C20})-(\ref{C22}), Eqs. (\ref{A14})-(\ref{A16}), and the conservation of the mean moment of inertia under tidal deformations, while $\delta E$ is obtained from Eqs. (\ref{C21}) and (\ref{A18}):
\begin{eqnarray}
 \label{A27}\delta C&=&M_e R^2 k_2\,e\,q_r \cos M+ k_2 \frac{4 R^5 n}{9 G} \dot \gamma,\\
 \label{A28}\delta E&=&-M_e R^2 k_2 q_r [i \sin{(\omega+M)}-\theta \sin{(\omega+M+\Omega-\psi)}].
\end{eqnarray}}
 
\subsection{Angular momentum equation}

The angular momentum equation can be written in an inertial frame, which is here taken to be a reference frame attached to the Laplace plane and centered at the center of mass of the satellite, as
\begin{equation}\label{A11}
\frac{d \vec L_{IN}}{dt}=\vec\Gamma_{IN},
\end{equation}
with $\vec L_{IN}$ the angular momentum of the satellite and $\vec\Gamma_{IN}$ the torque exerted on the satellite by the planet. To analytically study the spin precession, which is a long term-behavior, it is convenient to average this equation over the short orbital/diurnal period. 

The angular momentum in the inertial frame is obtained from the angular momentum in the BF thanks to the appropriate rotations
\begin{equation}\label{transL}
 \vec L_{IN}=R_z(-\psi).R_x(-\theta).R_z(-\phi).\vec L_{BF}.
\end{equation}
Neglecting wobble, and at first order, the angular momentum in the BF for the elastic case writes
\begin{equation}
\vec L_{BF}=\left(
\begin{array}{c}
-\delta E\, n \\
 0 \\
 C \,(n+\dot \gamma)+\delta C\, n
\end{array}\right),
\end{equation} 
where the spin precession rate $\dot \psi$ has been neglected in front of $n$ and where $\dot \gamma$ is dominated by the derivative of the diurnal librations ($\dot \gamma\varpropto \cos M$). \textcolor{black}{Using Eqs. (\ref{A27})-(\ref{A28}) for $\delta C$ and $\delta E$ and averaging over the orbital period, Eq. (\ref{transL}) becomes
\begin{eqnarray}
\vec L_{IN}&\simeq&\left(
\begin{array}{c}
 n\, C\, \theta\sin\psi\,  -\frac{n}{2}k_2 M_e R^2 q_r (\theta \sin\psi-i\sin\Omega)\\
 -n\, C\, \theta\cos\psi\, -\frac{n}{2}k_2 M_e R^2 q_r (-\theta \cos\psi+i\cos\Omega) \\
 n\, C 
 \end{array}\right)\\
\label{LIN}&\simeq& n\, C\, \hat s-\frac{n}{2}k_2 M_e R^2 q_r (\hat s-\hat n).
\end{eqnarray} 
with $\hat s=(s_x,s_y,s_z)$ and $\hat n=(n_x,n_y,n_z)$, the unit vectors along the spin axis and the normal to the orbit, expressed in the inertial frame:
\begin{eqnarray}
 (s_x,s_y)&\simeq&(\theta\cos{(\psi-\pi/2)},\theta\sin{(\psi-\pi/2)}),\\
 (n_x,n_y)&\simeq&(i\cos{(\Omega-\pi/2)},i\sin{(\Omega-\pi/2)}).
\end{eqnarray}
$\vec L_{IN}$ is only influenced by the obliquity tides ($\delta E$), since $\cos M=0$ (and so $\delta C=0$) on average over an orbital period. }

\textcolor{black}{The torque (\ref{A7}) is expressed in the inertial frame thanks to the appropriate rotations (the same as in Eq. (\ref{transL})) and is averaged over the orbital period with the slowly varying $\omega$, $\Omega$, and $\psi$ held constant:}
\begin{eqnarray}
\vec \Gamma_{IN}&\simeq&\left(
\begin{array}{c}
 \frac{3}{2} n^2 \left((C-A)-k_2 M_e R^2 q_r\right) (i \cos{\Omega}-\theta\cos{\psi})\\
 \frac{3}{2} n^2 \left((C-A)-k_2 M_e R^2 q_r\right) (i \sin{\Omega}-\theta\sin{\psi}) \\
 0\\
\end{array}\right)\\
\label{TIN}&=&\frac{3}{2} n^2 \lbrace(C-A)-k_2 M_e R^2 q_r\rbrace(\hat s \wedge \hat n).
\end{eqnarray}

\textcolor{black}{With the angular momentum (\ref{LIN}) and the torque (\ref{TIN}), the angular momentum equation (\ref{A11}) becomes}
\begin{equation}\label{AFinal}
n\, C \frac{d \hat s}{dt}-\frac{n}{2}k_2 M_e R^2 q_r \left(\frac{d \hat s}{dt}-\frac{d \hat n}{dt}\right)= \frac{3}{2} n^2 \left\lbrace(C-A)-k_2 M_e R^2 q_r\right\rbrace (\hat s \wedge \hat n),
\end{equation} 
which is equivalent to Eq. (\ref{elasticcase}). 

\section{Internal torque between two rigid layers separated by a liquid ocean}

We here provide a demonstration of Eq. (\ref{Kint}) for the strength of the internal gravitational torque, corrected for the pressure effect, between the rigid ice shell and the rigid interior separated by an internal liquid ocean. This demonstration aims to correct Eq. (30) of Baland et al. (2011) which has been used later in Baland et al. (2012) and Baland et al. (2014). 

\begin{figure}[!htb]
\begin{center}
\includegraphics[width=6.5cm]{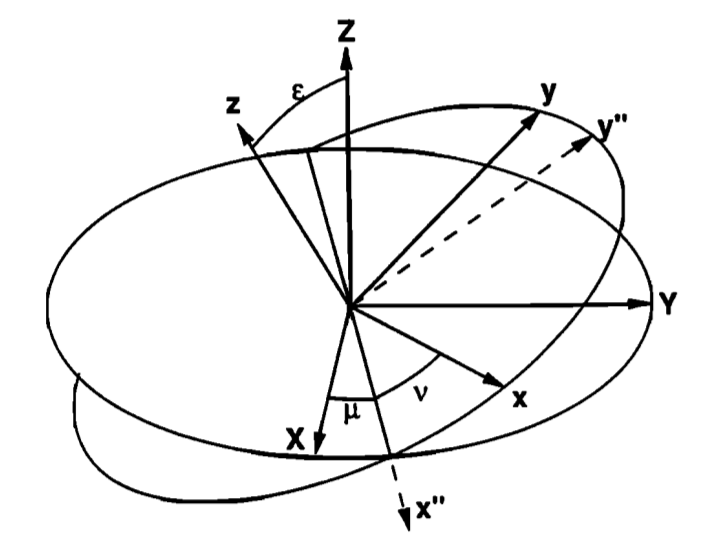}
\caption{\label{FigA3} Euler rotation angles between the shell and the interior Body Frames, denoted (X,Y,Z) and (x,y,z), respectively. This figure is reproduced from Szeto and Xu (1997). Note that here, $\varepsilon$ is the inclination between the two BFs, not to be confounded with $\varepsilon$, the \textcolor{black}{constant obliquity of a solid and rigid moon} defined in Section \ref{section2} of the present paper.}
\end{center}
\end{figure}

First, we forget the presence of the ocean, and we just consider the torque exerted by the shell on the interior. In Fig. 1 of Szeto and Xu (1997), reproduced here in Fig. \ref{FigA3}, the x axis (A-axis of the interior) and the X axis (A-axis of the shell) are about in the same direction, because of synchronous rotation. Therefore, the sum of Euler angles $\mu+\nu=0$ (note that $\mu$ is counted from the X axis), and the transformation matrix from the interior to the shell (Eq. (6) of Szeto and Xu 1997) reads:
\begin{eqnarray}
M_{in\mapsto sh}&=&R_z(-\mu).R_x(-\varepsilon).R_z(-\nu).
\end{eqnarray}
Using angles definition similar as in Fig. \ref{FigA2}, the subscripts $sh$ and $in$ corresponding to the solid shell and interior, the transformation matrix can be expressed in another way:
\begin{eqnarray}
 \nonumber M_{in\mapsto sh}&=&R_z(\phi_{sh}).R_x(\theta_{sh}).R_z(\psi_{sh}-\psi_{in}).R_x(-\theta_{in}).R_z(-\phi_{in})\\
 \label{Mshin}&=&\left( \begin{array}{ccc}
  l_1 & l_2 & l_3 \\
  m_1 & m_2 & m_3 \\
  n_1 & n_2 & n_3 \\
 \end{array}
 \right)=\left(
\begin{array}{c c c}
1 & l_2 & l_3 \\
-l_2 & 1 & m_3 \\
-l_3 & -m_3 & 1
\end{array}\right),
\end{eqnarray}
with
\begin{eqnarray}
\label{l2}l_2&=&\gamma_{sh}-\gamma_{in},\\
 l_3&=&\theta_{in} \sin{(M-\psi_{in}+\omega+\Omega)}- \theta_{sh} \sin{(M-\psi_{sh}+\omega+\Omega)},\\
\label{m3} m_3&=&\theta_{in} \cos{(M-\psi_{in}+\omega+\Omega)}- \theta_{sh} \cos{(M-\psi_{sh}+\omega+\Omega)}.
\end{eqnarray}
The components of the internal torque of the shell on the interior $\vec \Gamma_{int}$ can be computed from Eq. (7) of Szeto and Xu (1997):
\begin{eqnarray}
\nonumber \Gamma_{int}^X&=&\frac{4\pi G}{5}\left\lbrace(C_{in}-B_{in})\rho_{sh} (2\alpha_{sh}-2\alpha_o-\beta_{sh}+\beta_o)\right.\\
&&\left.\left[\theta_{in} \cos{(M-\psi_{in}+\omega+\Omega)}- \theta_{sh} \cos{(M-\psi_{sh}+\omega+\Omega)}\right]\right\rbrace,\\
\nonumber \Gamma_{int}^Y&=&\frac{4\pi G}{5}\left\lbrace-(C_{in}-A_{in})\rho_{sh} (2\alpha_{sh}-2\alpha_o+\beta_{sh}-\beta_o)\right.\\ 
&&\left.\left[\theta_{in} \sin{(M-\psi_{in}+\omega+\Omega)}- \theta_{sh} \sin{(M-\psi_{sh}+\omega+\Omega)}\right])\right\rbrace,\\
\Gamma_{int}^Z&=&\frac{4\pi G}{5}(B_{in}-A_{in})\rho_{sh} (\beta_{sh}-\beta_o)\sin 2(\gamma_{sh}-\gamma_{in}),
\end{eqnarray}
with $A_{in}<B_{in}<C_{in}$ the principal moments of inertia of the interior, $\alpha_{sh}$ and $\beta_{sh}$ the polar and equatorial flattenings of the shell,  and $\alpha_{o}$ and $\beta_{o}$ the polar and equatorial flattenings of the ocean. 

$\vec \Gamma_{int}$ is then transformed into the inertial frame with the following rotations\\ $R_z(-\psi_{in}).R_x(-\theta_{in}).R_z(-\phi_{in})$, and averaged over the orbital period, to give:
\begin{equation}
\label{B8} \vec \Gamma_{int}^{IN}=-\frac{8\pi G}{5}  M R^2 \left[-C_{20}^{in}\rho_{sh}(\alpha_{sh}-\alpha_o)+C_{22}^{in}\rho_{sh}(\beta_{sh}-\beta_o)\right] (\hat s_{sh} \wedge \hat s_{in}).
\end{equation}
$\hat s_{sh}$ and $\hat s_{in}$ are the unit vectors along the spin axis of the shell and of the interior, defined similarly as $\hat s$ in the solid case. The factors $C_{20}^{in}$ and $C_{22}^{in}$ represent the inertia of the interior, resisting to the gravity of the shell represented by the factors  $\rho_{sh}(\alpha_{sh}-\alpha_o)$ and $\rho_{sh}(\beta_{sh}-\beta_o)$. Equation (\ref{B8}) is easy to extend to the case with an ocean and to correct for the pressure torque. The top part of the ocean, which is aligned with the shell, also exerts a gravitational torque on the interior, and the ``shell factors'' have to be extended to the top ocean. The bottom ocean, which is aligned with the interior, exerts a pressure torque on the interior, which can be understood as a transfer to the interior of the gravitational torque exerted on the bottom ocean by the shell and the top ocean. Therefore, the ``interior factors'' have to be extended to the bottom ocean. We obtain 
\begin{eqnarray} \nonumber\vec \Gamma_{int}^{IN}&=&K n (\hat s_{sh} \wedge \hat s_{in})\\
\nonumber K&=&-\frac{8\pi G}{5 n}  M_e R^2 \left[-(C_{20}^{in}+C_{20}^{b,o})[\rho_{sh}(\alpha_{sh}-\alpha_o)+\rho_o \alpha_o]\right.\\
\label{B10}&&\left. +(C_{22}^{in}+C_{22}^{b,o})[\rho_{sh} (\beta_{sh}-\beta_o)+\rho_o\beta_o]\right].
\end{eqnarray}
The last equation is equivalent to Eq. (\ref{Kint}). 

In Baland et al. (2011), we made the additional but incorrect assumption that $\mu=-\pi/2$ and $\nu=\pi/2$, based on the incorrect intuition that the B axes of the solid layers have to be exactly aligned with each other in the Cassini state. However, the layers are rotating and $\mu$ and $\nu$ do not have a fixed value. Applied to Enceladus, this error would lead to overestimate the numerical value of $K$ by a factor 2. We show in Section \ref{computation} that the shell/interior obliquity is smaller/larger than the \textcolor{black}{obliquity of a solid Enceladus}. Therefore, it is understandable that such an overestimation of the internal coupling strength would lead to bring the obliquity of the solid layers closer to each other, that is to say to overestimate/underestimate the shell/interior obliquity by about $20\%$ for Enceladus. This error has consequences on the predictions made in Baland et al. (2012) for the obliquity of the Galilean satellites. For instance, the range for the shell obliquity of Europa has in fact to be lower than the range [$0.033^\circ$,$0.044^\circ]$ given in Table 6 of Baland et al. (2012). For Titan, this error has lead Baland et al. (2014) to underestimate the free period $T_+$ \textcolor{black}{by about 25\%, and to conclude that it was smaller than the main period of the orbital precession, while it can be smaller or larger}. As a result, they found fewer density profiles with the appropriate resonant amplification to account for the measured obliquity than with the corrected model \textcolor{black}{(233 profiles instead of 1979, out of the 905\,014 profiles considered at the beginning). However, these 1979 profiles are generally similar to the 233 profiles found in the first place, and the conclusions drawn for the constraints on Titan's interior are not much affected. For instance, considering also the information brought by the measured Love number $k_2$ (see seventh column of Table 4 in Baland et al., 2014), we now obtain a normalized mean moment of inertia between 0.30 and 0.33 (instead of 0.31 and 0.32, respectively). This range is still lower than the expected hydrostatic value (0.34).} 

\section{Additional internal torques between two elastic layers separated by a liquid ocean}

We here provide a demonstration of Eqs. (\ref{Kpi})-(\ref{Kps}) for the strengths $K_{pi}$ and $K_{ps}$ of the internal gravitational torque, corrected for the pressure effect, between the static bulge of the ice shell/interior and the periodic tidal obliquity bulge of the interior/ice shell. This demonstration consists in an extension of the computation presented in Szeto and Xu (1997) for the torque exerted by the external rigid ellipsoidal shell on the rigid interior. 

\subsection{Gravitational potential in the cavity created by the shell}

Replacing equation (4) of Szeto and Xu (1997), deformation from a sphere of any interface between two layers reads
\begin{eqnarray}
\nonumber r&=&R_j\left(1-\frac{2}{3}\alpha_j R_{20}-\frac{2}{3}\delta\alpha^{rad}_j R_{20}-\frac{2}{3}\delta\alpha^{obl}_jR_{21}+\frac{1}{6}\beta_jR_{22}^{cos}\right.\\
\label{rj}&&\left.+\frac{1}{6}\delta\beta^{rad}_jR_{22}^{cos}+\frac{1}{6}\delta\beta^{lib}_jR_{22}^{sin}\right),
\end{eqnarray}
where $R_j$ is the mean radius of the interface and
\begin{eqnarray}
R_{20}&=&\frac{1}{2} \left(-1+3 \cos^2\varphi\right),\\
R_{21}&=&-\frac{3}{2} \sin2\varphi\, \cos\lambda,\\ 
R_{22}^{cos}&=&3\sin^2\varphi\cos2\lambda, \\
R_{22}^{sin}&=&3\sin^2\varphi\sin2\lambda,
\end{eqnarray}
with the colatitude and longitude denoted by $\varphi$ and $\lambda$, respectively. The polar and equatorial flattenings associated with the static bulge are denoted by $\alpha_j$ and $\beta_j$. The periodic flattenings associated with the periodic tidal bulges of fixed orientations and varying amplitudes (see Appendix A for the definition of the different kinds of bulge) are denoted by $\delta\alpha^{rad}_j$ and $\delta\beta^{rad}_j$ for the radial bulge, by $\delta\beta^{lib}_j$ for the librational bulge, and by $\delta\alpha^{obl}_j$ for the obliquity bulge. \textcolor{black}{The use of $\delta$ indicates that these periodic flattenings are an order smaller in the small quantities $e$, $\gamma$, and $i$ and $\theta$ than the static flattenings.}

Since the periodic tidal bulges are an order \textcolor{black}{smaller in $e$, $\gamma$, and $i$ and $\theta$ than the static bulges}, we can neglect the small misalignment between the different layers when computing the additional elastic torques. We also neglect the torques between the small radial bulges and the small librational and obliquity bulges, to keep only ``static-periodic'' torques. The periodic radial flattenings can therefore be neglected, and Eq. (\ref{rj}) can be simplified to   
\begin{eqnarray}
\label{rj2} r&=&R_j\left(1-\frac{2}{3}\alpha_j R_{20}-\frac{2}{3}\delta\alpha^{obl}_jR_{21}+\frac{1}{6}\beta_jR_{22}^{cos}+\frac{1}{6}\delta\beta^{lib}_jR_{22}^{sin}\right).
\end{eqnarray}

The gravitational potential in the cavity created by the shell, $\Phi$, is expressed by Eq. (A2) of Szeto and Xu (1997), in function of some integrals denoted $I_{n,m}$ defined by their equation (A3). Using equation (\ref{rj2}) of the present paper instead of equation (4) of Szeto and Xu (1997), in their equations (A2-A3), we find
\begin{eqnarray}
 \Phi&=&\phi_{20}r^2 R_{20} +\delta\phi_{21}^{obl} r^2 R_{21} + \phi_{22}r^2 R_{22}^{cos} +\delta\phi_{22}^{lib} r^2 R_{22}^{sin},
 \end{eqnarray}
with
\begin{eqnarray}
 \label{phi20}\phi_{20}&=&-\frac{8\pi G}{15}\rho_{sh} (\alpha_{sh}-\alpha_o),\\
 \delta \phi_{21}^{obl}&=&-\frac{8\pi G}{15}\rho_{sh} (\delta\alpha_{sh}^{obl}-\delta\alpha_o^{obl}),\\
 \label{phi22}\phi_{22}&=&\frac{2\pi G}{15}\rho_{sh} (\beta_{sh}-\beta_o),\\
 \label{phi22lib}\delta\phi_{22}^{lib}&=&\frac{2\pi G}{15}\rho_{sh}(\delta\beta_{sh}^{lib}-\delta\beta_o^{lib}).
\end{eqnarray}

The coefficients $\delta\phi_{21}^{obl}$ and $\delta\phi_{22}^{lib}$ depend on the periodic flattenings $\delta\alpha^{obl}_j$ and $\delta\beta^{lib}_j$ of a layer $j$, due to obliquity and librational tides. Using Eq. (\ref{rj2}), these flattenings can be related to the radial displacements of the outer surface of the layer due to obliquity and librational tides, as
\begin{eqnarray}
 \xi^{obl}_{j}&=&\frac{-2}{3}R_j\delta\alpha^{obl}_j R_{21}, \\
 \xi^{lib}_{j}&=&\frac{1}{6}R_j\delta\beta^{lib}_j R_{22}^{sin}.
\end{eqnarray}
The radial displacements $\xi^{obl}_{j}$ and $\xi^{lib}_{j}$ can also be defined in terms of the radial tidal displacement $y_j$ at radius $R_j$, calculated for a tidal potential with degree-two component equal to $-1$ m$^2/$s$^2$ at the surface:
\begin{eqnarray}
 \xi^{obl}_{j}&=&-y_j  V_t^{obl}(R), \\
 \xi^{lib}_{j}&=&-y_j V_t^{lib}(R),
\end{eqnarray}
where $V_t^{obl}(R)$ and $V_t^{lib}(R)$ are the part of the tidal potential (\ref{A4}) inducing the obliquity and librational bulges evaluated at $R$, respectively:
\begin{eqnarray}
 V_t^{obl}(R)&=& n^2 R^2 [\theta_j \sin(\omega+M+\Omega-\psi_j)-i \sin(\omega+M)- ] R_{21},\\
 V_t^{lib}(R)&=&-n^2 R^2 (e \sin M-\gamma_j/2) R_{22}^{sin}. 
\end{eqnarray}
Therefore
\begin{eqnarray}
\delta\alpha^{obl}_j&=&\frac{3}{2}\frac{y_j}{R_j} n^2 R^2 [\theta_j \sin(\omega+M+\Omega-\psi_j)-i \sin(\omega+M)], \\
 \delta\beta^{lib}_j   &=&6 \frac{y_j}{R_j} n^2 R^2 (e \sin M-\gamma_j/2), 
\end{eqnarray}
and
\begin{eqnarray}
\nonumber\delta \phi_{21}^{obl}&=&-\frac{4\pi G}{5} n^2 R^2\rho_{sh} \left[\frac{y_{sh}}{R} -\frac{y_o}{R_o} \right]\\
&&[\theta_{sh} \sin(\omega+M+\Omega-\psi_{sh})-i \sin(\omega+M)], \\
\delta\phi_{22}^{lib}&=&\frac{4\pi G}{5}  n^2 R^2\rho_{sh} \left[ \frac{y_{sh}}{R} - \frac{y_o}{R_o} \right] (e \sin M-\gamma_{sh}/2).
\end{eqnarray}

\subsection{Additional torques on the interior}

Since we neglect the effect of radial periodic bulges, we only need to consider for the interior the principal moments of inertia $A_{in}, B_{in}$ and $C_{in}$, representing the static bulge, and the additional moments of inertia $\delta E_{in}$ and $\delta F_{in}$, representing the obliquity and librational periodic bulges of the interior. They are defined similarly as in the solid case (\ref{A14}-\ref{A19}) by:
\begin{eqnarray}
 (C_{in}-A_{in})&=&M_e R^2 (-C_{20}^{in}+2C_{22}^{in}),\\
 (C_{in}-B_{in})&=&M_e R^2 (-C_{20}^{in}-2C_{22}^{in}),\\
 (B_{in}-A_{in})&=&4 M_e R^2 C_{22}^{in} ,\\ 
 \delta E_{in}&=&-M_e R^2 C_{21}^{in},\\
 \delta F_{in}&=& 2 M_e R^2 S_{22}^{in}.
\end{eqnarray}
The additional torque (an extension of Eq (7) of Szeto and Xu, 1997) is:
\begin{eqnarray}
\nonumber\delta \vec \Gamma_{int} &=&3 \phi_{20}  \left(
\begin{array}{c}
 -n_1 n_2 \delta E_{in}+ n_1 n_3 \delta F_{in}\\
(n_1^2-n_3^2) \delta E_{in} -n_2 n_3 \delta F_{in} \\
n_2 n_3\delta E_{in}+  (-n_1^2+n_2^2) \delta F_{in} \\
\end{array}
\right)\\
\nonumber &&+6\phi_{22}  \left(
\begin{array}{c}
(-l_1 l_2 + m_1 m_2) \delta E_{in}+(l_1 l_3 - m_1 m_3)\delta F_{in}\\
(l_1^2 - l_3^2 - m_1^2 + m_3^2) \delta E_{in}+ (-l_2 l_3 + m_2 m_3)\delta F_{in}\\
(l_2 l_3 - m_2 m_3) \delta E_{in}+(-l_1^2 + l_2^2 + m_1^2 - m_2^2) \delta F_{in}\\
\end{array}
\right)\\
\nonumber &&+6\delta \phi_{22}^{lib}  \left(
\begin{array}{c}
 (l_3 m_2 + l_2 m_3)(C_{in}-B_{in})\\
 -(l_3 m_1 + l_1 m_3)(C_{in}-A_{in})\\
 (l_2 m_1 + l_1 m_2) (B_{in}-A_{in})\\
\end{array}
\right)\\
&&\label{deltatorque}+3\delta \phi_{21}^{obl}  \left(
\begin{array}{c}
 - (l_3 n_2 + l_2 n_3) (C_{in}-B_{in})\\
 (l_3 n_1 + l_1 n_3) (C_{in}-A_{in})\\
  -(l_2 n_1 + l_1 n_2) (B_{in}-A_{in})\\
\end{array}
\right),
\end{eqnarray}
with $l_{i}, m_{i}$ and $n_{i}$ the components of the transformation matrix from the interior to the shell defined by Eq. (\ref{Mshin}). Using Eqs. (\ref{l2})-(\ref{m3}), the additional torque becomes
\begin{eqnarray}
\delta \vec \Gamma_{int} &=& \left(
\begin{array}{c}
0\\
-3\delta E_{in} (\phi_{20}-2\phi_{22}) +3(C_{in}-A_{in}) \delta \phi_{21}^{obl}\\
  -12 \phi_{22}\delta F_{in}+6\delta\phi_{22}^{lib}(B_{in}-A_{in}) \\
\end{array}
\right).
\end{eqnarray}

The z component of $\delta \vec \Gamma_{int} $ is the part of the additional torque that induce longitudinal librations (see Eq. (B13) of Van Hoolst et al., 2013) and the y-component is implicated in the spin precession. The first part of the y component is for the torque of the static bulge of the shell on the periodic bulges of the interior (first two lines in Eq. (\ref{deltatorque})) and the second part is for the torque of the periodic bulges on the static bulge of the interior (last two lines in Eq. (\ref{deltatorque})). 

Now this torque has to be transformed into the inertial frame with the following rotations $R_z(-\psi_{in}).R_x(-\theta_{in}).R_z(-\phi_{in})$, and averaged over the orbital period, to give:
\begin{eqnarray}\nonumber \delta\vec \Gamma_{int}^{IN}
&=&\frac{3}{2}k_2^{in} M R^2 q_r (\phi_{20}-2\phi_{22}) (\hat n \wedge \hat s_{in})\\
\label{C40}&+&\frac{6\pi G}{5}n^2 R^2 \rho_{sh} \left[\frac{y_{sh}}{R} -\frac{y_o}{R_o} \right](C_{in}-A_{in}) (\hat n \wedge \hat s_{sh}),
\end{eqnarray}
where the effects of librational tides disappear during the averaging, only leaving the effect of the obliquity tides. 

In the libration study (Van Hoolst et al., 2013), the additional elastic torques were not due, at first order, to misalignment between layers. The librations angle of a layer $\gamma_j$ appeared in the torques only because $\gamma_j$ modulates the amplitudes of the librational bulge of layer $j$. Similarly, the spin inclination of a given layer $\theta_j$ appears only (through $\hat s_{j}$) because of its role in the modulation of the amplitude of the obliquity bulge of layer $j$. Neglecting misalignment between layers, the obliquity bulge of a layer is symmetric with respect to the $(C,A)$ plane of the layer, and misaligned by $\pi/4$ with respect to the static bulge. Therefore the torque between the periodic bulge of layer $j$ and the static bulge of the other solid layer is perpendicular to the $(C,A)$ plane, which is equivalent, at first order to the plane defined by the vectors $\hat n$ and $\hat s_{j}$.

Equation (\ref{C40}) for the additional elastic torques can be extended to the ocean and pressure effects, just as for Eq. (\ref{B10}) for the rigid internal torque:

\begin{eqnarray} 
\label{C30}\delta\vec \Gamma_{int}^{IN}&=&n K_{pi} (\hat n \wedge \hat s_{in})+n K_{ps} (\hat n \wedge \hat s_{sh}),\\
K_{pi}&=&\frac{3}{2 n}(k_2^{in}+k_2^{o,b}) M R^2 q_r (\phi_{20}-2\phi_{22}+\phi_{20}^{o,t}-2\phi_{22}^{o,t}),\\
\nonumber K_{ps}&=&\frac{6\pi G}{5}n R^2 \left[\rho_{sh} \left(\frac{y_{sh}}{R} -\frac{y_o}{R_o} \right)+\rho_o\frac{y_o}{R_o} \right]\\
 &&[(C_{in}-A_{in})+(C_{o,b}-A_{o,b})].
\end{eqnarray}
This completes the demonstration of Eqs. (\ref{Kpi})-(\ref{Kps}).


\newpage
\begin{thebibliography}{}

 
\bibitem{1} Baland, R.-M., Van Hoolst, T., Yseboodt, M., Karatekin, \"O., 2011. Titan's obliquity as evidence of a subsurface ocean? Astronomy and Astrophysics, 530, A141+.
 
\bibitem{2} Baland, R.-M., Yseboodt, M., Van Hoolst, T., 2012. Obliquity of the Galilean satellites: The influence of a global internal liquid layer. Icarus, Volume 220, Issue 2, p. 435-448. 
 
\bibitem{3} {{Baland}, R.-M., {Tobie}, G., {Lef{\`e}vre}, A., {Van Hoolst}, T.}, 2014. Titan's internal structure inferred from its gravity field, shape, and rotation state. Icarus, Volume 237, p. 29-41.

\bibitem{4} {{Beuthe}, M.}, \textcolor{black}{2013}. Spatial patterns of tidal heating. Icarus, Volume 223, p. 308-329. 

\bibitem{5} Bills, B.G., 2005. Free and forced obliquities of the Galilean satellites of Jupiter. Icarus, Volume 175, p. 233-247.
  
\bibitem{6} Bills, B.G., Nimmo, F., 2008. Forced obliquity and moments of inertia of Titan. Icarus 196, 293-297.

\bibitem{7} Chen, E.M.A, Nimmo, F., 2011. Obliquity tides do not significantly heat Enceladus, Icarus, Volume. 214, p. 779-781.

\bibitem{8} Coyette, A., Van Hoolst, T., Baland, R.-M., Tokano, T., 2016. Modeling the polar motion of Titan, \textcolor{black}{Icarus, doi:10.1016/j.icarus.2015.10.015.}

\bibitem{9} Dahlen, F.A., Tromp, J., 1999. Theoretical Global Seismology. Princeton University Press, Princeton.

\bibitem{10} {Gastineau, Micka\"{e}l and Laskar, Jacques}, 2011. TRIP: A Computer Algebra System Dedicated to Celestial Mechanics and Perturbation Series, ACM Commun. Comput. Algebra, Volume. 44, p. 194--197.

\bibitem{11}{{Giese}, B., {Hussmann}, H., {Roatsch}, T., {Helfenstein}, P., 
	{Thomas}, P.~C., {Neukum}, G.}, 2011. Enceladus: Evidence for librations forced by Dione. EPSC-DPS Joint Meeting 2011, p 976.

\bibitem{12}{{Giese}, B.}, 2014. An upper limit on Enceladus obliquity. EPSC 2014, p. 419.


\bibitem{13} \textcolor{black}{{{Grasset}, O. and {Dougherty}, M.~K. and {Coustenis}, A. and 
	{Bunce}, E.~J. and {Erd}, C. and {Titov}, D. and {Blanc}, M. and 
	{Coates}, A. and {Drossart}, P. and {Fletcher}, L.~N. and {Hussmann}, H. and 
	{Jaumann}, R. and {Krupp}, N. and {Lebreton}, J.-P. and {Prieto-Ballesteros}, O. and 
	{Tortora}, P. and {Tosi}, F. and {Van Hoolst}, T.}, 2013. JUpiter ICy moons Explorer (JUICE): An ESA mission to orbit Ganymede and to characterise the Jupiter system, Planetary and Space Science, Volume 78, p. 1-21. }
	
\bibitem{14}  \textcolor{black}{{Henrard}, J., 2005. The Rotation of Europa, Celestial Mechanics and Dynamical Astronomy, Volume 91, p. 131-149. }
	

\bibitem{15}{{Iess}, L., {Stevenson}, D.~J., {Parisi}, M., {Hemingway}, D., 
	{Jacobson}, R.~A., {Lunine}, J.~I., {Nimmo}, F., {Armstrong}, J.~W., 
	{Asmar}, S.~W., {Ducci}, M., {Tortora}, P.}, 2014. The Gravity Field and Interior Structure of Enceladus. Science, Vol 344, p. 78-80.

\bibitem{16} {{Jara-Oru{\'e}}, H.~M., {Vermeersen}, B.~L.~A.}, 2014. The forced libration of Europa's deformable shell and its dependence on interior parameters, Icarus, Volume 229, p. 31-44. 

\bibitem{17} {{Lainey}, V., {Arlot}, J.-E., {Karatekin}, {\"O}., {Van Hoolst}, T.}, 2009. Strong tidal dissipation in Io and Jupiter from astrometric observations, Nature, Volume 459, p. 957-959. 

\bibitem{18} {{McKinnon}, W.~B.}, 2015. \textcolor{black}{Effect of Enceladus's rapid synchronous spin on interpretation of Cassini gravity}, Geophysical Research Letters, Volume 42, Issue 7, pp. 2137-2143. 

\bibitem{19} Moore, W. B., Schubert, G., 2000. The tidal response of Europa. Icarus, 147, 317-319.

\bibitem{20} Moore, W. B., Schubert, G., 2003. The tidal response of Ganymede and Callisto with and without liquid water oceans. Icarus, 166, 223-226.

\bibitem{21} Murray, C.D., Dermott, S.F., 1999. Solar System Dynamics. Cambridge University Press.

\bibitem{22} {{Nimmo}, F. and {Bills}, B.~G. and {Thomas}, P.~C.}, 2011. Geophysical implications of the long-wavelength topography of the Saturnian satellites. Journal of Geophysical Research, Volume 116, Issue E11, CiteID E11001.

\bibitem{23} Nimmo, F., Porco, C., Mitchell, C., 2014. Tidally Modulated Eruptions on Enceladus: Cassini ISS Observations and Models. The Astronomical Journal, Volume 148, Issue 3, article id. 46, 14 pp. 

\bibitem{24} {{Nimmo}, F. and {Spencer}, J.~R.}, 2015. Powering Triton's recent geological activity by obliquity tides: Implications for Pluto geology. Icarus, 246, 2-10.

\bibitem{25} Noyelles., B., 2010. Theory of the rotation of Janus and Epimetheus. Icarus 207, 887-902.

\bibitem{26}  \textcolor{black}{{Patthoff}, D.~A. and {Kattenhorn}, S.~A.., 2011. A fracture history on Enceladus provides evidence for a global ocean.
	Geophysical Research Letters, Volume 38, Issue 18, CiteID L18201}

\bibitem{27} Peale., S.~J., 1969. Generalized Cassini's Laws. Astronomical Journal 74, 483-489.

\bibitem{28}{{Porco}, C.~C., {Helfenstein}, P., {Thomas}, P.~C., {Ingersoll}, A.~P., {Wisdom}, J., {West}, R., {Neukum}, G., {Denk}, T., {Wagner}, R., {Roatsch}, T.,{Kieffer}, S., {Turtle}, E., {McEwen}, A., {Johnson}, T.~V., {Rathbun}, J., {Veverka}, J., {Wilson}, D., {Perry}, J., {Spitale}, J., {Brahic}, A., {Burns}, J.~A., {Del Genio}, A.~D., {Dones}, L., {Murray}, C.~D., {Squyres}, S.}, 2006. Cassini Observes the Active South Pole of Enceladus. Science, Volume 311, p. 1393-1401.
  
\bibitem{29}  \textcolor{black}{{{Postberg}, F. and {Schmidt}, J. and {Hillier}, J. and {Kempf}, S. and 
	{Srama}, R.}, 2011. A salt-water reservoir as the source of a compositionally stratified plume on Enceladus. Nature, Vol 474 253, p. 620-622.}
	
\bibitem{30} \textcolor{black}{{{Rambaux}, N. and {Castillo-Rogez}, J.~C. and {Williams}, J.~G. and {Karatekin}, {\"O}.}, 2010. Librational response of Enceladus. Geophysical Research Letters, Volume 37, Issue 4, CiteID L04202.}
	
\bibitem{31} {{Rhoden}, A.~R. and {Hurford}, T.~A. and {Roth}, L. and {Retherford}, K.}, 2015. Linking Europa's plume activity to tides, tectonics, and liquid water. Icarus 253, 169-178.

\bibitem{32} {{Richard}, A., {Rambaux}, N., {Charnay}, B.}, 2014. Librational response of a deformed 3-layer Titan perturbed by non-Keplerian orbit and atmospheric couplings, Planetary and Space Science, Volume 93, p. 22-34.

\bibitem{33} \textcolor{black}{{{Rivoldini}, A. and {Van Hoolst}, T. and {Verhoeven}, O.}, 2009. The interior structure of Mercury and its core sulfur content. Icarus 201, 12-30.}

\bibitem{34}  \textcolor{black}{{{Roberts}, J.~H. and {Nimmo}, F.}, 2008. Tidal heating and the long-term stability of a subsurface ocean on Enceladus. Icarus 194, 675-689.}

\bibitem{35} \textcolor{black}{{{Spencer}, J.~R. and {Pearl}, J.~C. and {Segura}, M. and {Flasar}, F.~M. and 
	{Mamoutkine}, A. and {Romani}, P. and {Buratti}, B.~J. and {Hendrix}, A.~R. and 
	{Spilker}, L.~J. and {Lopes}, R.~M.~C.}, 2006. Cassini Encounters Enceladus: Background and the Discovery of a South Polar Hot Spot. Science, Vol 311, p. 1401-1405.}
	
\bibitem{36} \textcolor{black}{{{Spencer}, J.~R. and {Barr}, A.~C. and {Esposito}, L.~W. and 
	{Helfenstein}, P. and {Ingersoll}, A.~P. and {Jaumann}, R. and 
	{McKay}, C.~P. and {Nimmo}, F. and {Waite}, J.~H.}, 2009. Enceladus: An Active Cryovolcanic Satellite. In \textit{Saturn from Cassini-Huygens}, pages 683-724. Springer. }	

\bibitem{37} \textcolor{black}{Sotin, C., Grasset, O., Beauchesne, S., 1998. Thermodynamical properties of high
pressure ices: Implications for the dynamics and internal structure of large icy
satellites. In: Schmitt, B., De Bergh, M., Festou, C. (Eds.), Solar System Ices.
Kluwer Academic, Dordrecht, The Netherlands, pp. 79–96. }	

\bibitem{38} Stiles, B.W., Kirk, R.L., Lorenz, R.D., Hensley, S., Lee, E., Ostro, S.J., Allison, M.D., Callahan, P.S., Gim, Y., Iess, L., Persi del Marmo, P., Hamilton, G., Johnson, W.T.K., West, R.D., 2008. Determining Titan's Spin State from Cassini Radar Images. Astron. J. 135, p. 1669-1680.
  
\bibitem{39} Stiles, B.W., Kirk, R.L., Lorenz, R.D., Hensley, S., Lee, E., Ostro, S.J., Allison, M.D., Callahan, P.S., Gim, Y., Iess, L., Persi del Marmo, P., Hamilton, G., Johnson, W.T.K., West, R.D., 2010. ERRATUM: ''Determining Titan's Spin State from Cassini Radar Images''. Astron. J. 139, p. 311. 

\bibitem{40}  {{Szeto}, A.~M.~K., {Xu}, S.}, 1997. Gravitational coupling in a triaxial ellipsoidal Earth. Journal of Geophysical Research, Volume 102, Issue B12, p. 27651-27658

\bibitem{41} \textcolor{black}{{Thomas}, P.~C. and {Tajeddine}, R. and {Tiscareno}, M.~S. and 
	{Burns}, J.~A. and {Joseph}, J. and {Loredo}, T.~J. and {Helfenstein}, P. and 
	{Porco}, C., 2016. Enceladus's measured physical libration requires a global subsurface ocean. Icarus, 264, 37-47. }

\bibitem{42} Tobie, G., Mocquet, A., Sotin, C., 2005. Tidal dissipation within large icy satellites: Applications to Europa and Titan. Icarus, 177, 534-549. 

\bibitem{43} \textcolor{black}{{{Tobie}, G. and {{\v C}adek}, O. and {Sotin}, C.}, 2008. Solid tidal friction above a liquid water reservoir as the origin of the south pole hotspot on Enceladus. Icarus, 196, 642-652.}

\bibitem{44} \textcolor{black}{Turcotte, D.L., Schubert, G., 2002. Geodynamics. Cambridge University Press.}

\bibitem{45} {{Tyler}, R.~H.}, 2009. Ocean tides heat Enceladus. Geophysical Research Letters, vol. 36, issue 15, p. L15205.

\bibitem{46} {{Tyler}, R.~H.}, 2011. Tidal dynamical considerations constrain the state of an ocean on Enceladus. Icarus, Volume 211, Issue 1, p. 770-779.

\bibitem{47} Van Hoolst, T., Rambaux, N., Karatekin, \"O., Dehant, V., Rivoldini, A., 2008. The librations, shape, and icy shell of Europa. Icarus 195/1, 386-399.

\bibitem{48} Van Hoolst, T., Rambaux, N.,  Karatekin, \"O., Baland, R.-M., 2009. The effect of gravitational and pressure torques on Titan's length-of-day variations. Icarus, 200, 256-264.

\bibitem{49} Van Hoolst, T., Baland, R.-M., Trinh, A., 2013. The effect of tides on the longitudinal librations of large synchronously rotating icy satellites. Icarus, 226, 299-315.

\bibitem{50} Vienne, A., Duriez, L., 1995. TASS1.6: Ephemerides of the major Saturnian satellites. Astronomy and Astrophysics, v.297, p.588.

\bibitem{51} {{Wahr}, J.~M. and {Zuber}, M.~T. and {Smith}, D.~E. and {Lunine}, J.~I.}, 2006. Tides on Europa, and the thickness of Europa's icy shell. Journal of Geophysical Research (Planets), Volume 111, Issue E12, CiteID E12005.
\end{thebibliography}
\end{document}